\documentclass[10pt,twocolumn,letterpaper]{article}

\usepackage[pagenumbers]{cvpr} 

\definecolor{cvprblue}{rgb}{0.21,0.49,0.74}
\usepackage[pagebackref,breaklinks,colorlinks,allcolors=cvprblue]{hyperref}

\def\paperID{*****} 
\def\confName{CVPR}
\def\confYear{2025}

\title{Learning Image-Text Matching with Optimal Partial Transport}

\author{
Zhengxin Pan\\
Zhejiang University\\
{\tt\small panzx@zju.edu.cn}
\and
Haishuai Wang\\
Zhejiang University\\
{\tt\small haishuai.wang@zju.edu.cn}
\and
Fangyu Wu\\
Xi'an Jiaotong-Liverpool University\\
{\tt\small Fangyu.wu02@xjtlu.edu.cn}
\and
Bailing Zhang\\
NingboTech University\\
{\tt\small bailing.zhang@nit.zju.edu.cn}
\and
Jiajun Bu\\
Zhejiang University\\
{\tt\small bjj@zju.edu.cn}
\and
Hongyang Chen\\
Zhejiang Lab\\
{\tt\small dr.h.chen@ieee.org}
}

\usepackage{booktabs}
\usepackage{calc}
\usepackage{multicol}
\usepackage{multirow}%
\usepackage{color}
\usepackage{siunitx}
\usepackage{subcaption} %

\usepackage{algorithm}
\usepackage{algorithmic}

\usepackage{tocloft}

\newcommand{\first}[1]{\textcolor{red}{\bf#1}}
\newcommand{\cmmnt}[1]{}

\usepackage{amssymb}
\usepackage{amsmath,amsfonts}
\usepackage{amsopn,bm}

\usepackage{nicefrac}       

\usepackage{epsfig}
\usepackage{graphicx}
\usepackage{float}

\usepackage{bm,xspace}
\usepackage{comment}
\usepackage{verbatim}
\usepackage{multirow}
\usepackage{makecell}
\usepackage{array}
\usepackage{url}
\usepackage{booktabs}
\usepackage{etoolbox}
\usepackage{calc}
\usepackage{pifont,hologo}
 
\usepackage{nicefrac}

\usepackage{arydshln}
\usepackage[utf8]{inputenc} 
\usepackage[T1]{fontenc}    
\usepackage{url}            
\usepackage{amsfonts}       
\usepackage{nicefrac}       
\usepackage{microtype}      
\usepackage[american]{babel}
\usepackage{enumitem}
\usepackage{siunitx}

\usepackage{enumitem}

\usepackage[super]{nth}
\usepackage{nicefrac}
\robustify\bfseries

\usepackage{dsfont}

\usepackage{listings}
\usepackage{xcolor}
\definecolor{codegreen}{rgb}{0,0.6,0}
\definecolor{codegray}{rgb}{0.5,0.5,0.5}
\definecolor{codepurple}{rgb}{0.58,0,0.82}
\definecolor{backcolour}{rgb}{1.0,1.0,1.0}

\lstdefinestyle{mystyle}{
    commentstyle=\color{codegreen},
    keywordstyle=\color{magenta},
    numberstyle=\tiny\color{codegray},
    stringstyle=\color{codepurple},
    basicstyle=\ttfamily\footnotesize,
    breakatwhitespace=false,         
    breaklines=true,                 
    captionpos=b,                    
    keepspaces=true,                 
    numbersep=0pt,                  
    showspaces=false,                
    showstringspaces=false,
    showtabs=false,                  
    tabsize=2,
    linewidth=.99\textwidth,
    xleftmargin=0.01cm
}
\usepackage{mathtools}

\newcommand{\vct}[1]{\boldsymbol{#1}} 
\newcommand{\mat}[1]{\boldsymbol{#1}} 

\newcommand{\field}[1]{\mathbb{#1}}
\newcommand{\R}{\field{R}} 

\newcommand{\ProbOpr}[1]{\mathbb{#1}}

\newcommand{\expect}[2]{%
\ifthenelse{\equal{#2}{}}{\ProbOpr{E}_{#1}}
{\ifthenelse{\equal{#1}{}}{\ProbOpr{E}\left[#2\right]}{\ProbOpr{E}_{#1}\left[#2\right]}}} 
\newcommand{\var}[2]{%
\ifthenelse{\equal{#2}{}}{\ProbOpr{VAR}_{#1}}
{\ifthenelse{\equal{#1}{}}{\ProbOpr{VAR}\left[#2\right]}{\ProbOpr{VAR}_{#1}\left[#2\right]}}} 

\newcommand{\ones}{\vct{1}}

\newcommand{\vt}{\vct{t}}

\newcommand{\vv}{\vct{v}}

\newcommand{\mC}{\mat{C}}

\newcommand{\mK}{\mat{K}}

\newcommand{\mU}{\mat{U}}

\newcommand{\vtheta}{\vct{\theta}}
\newcommand{\vmu}{\vct{\mu}}
\newcommand{\mSigma}{\mat{\Sigma}}
\newcommand{\mOmega}{\mat{\Omega}}

\newcommand{\mxi}{\mat{\xi}}

\newcommand{\valpha}{\vct{\alpha}}
\newcommand{\vbeta}{\vct{\beta}}

\newcommand{\vgamma}{\vct{\gamma}}
\newcommand{\mGamma}{\mat{\Gamma}}

\def\calL{\mathcal{L}}

\def\calT{\mathcal{T}}

\def\calV{\mathcal{V}}

\makeatletter
\DeclareRobustCommand\onedot{\futurelet\@let@token\@onedot}
\def\@onedot{\ifx\@let@token.\else.\null\fi\xspace}

\makeatother

\newcommand{\symtext}[2]{\textsc{#1}#2}
\newcommand{\eat}[1]{{}}

\newcommand\mypara[1]{\vspace{1mm}\noindent\textbf{#1}}

\usepackage{array}
\usepackage{xcolor, colortbl}

\definecolor{Gray}{gray}{0.5}

\newlength\savewidth

\makeatletter\renewcommand\paragraph{\@startsection{paragraph}{4}{\z@}
  {.5em \@plus1ex \@minus.2ex}{-.5em}{\normalfont\normalsize\bfseries}}\makeatother

\newcolumntype{x}[1]{>{\centering\arraybackslash}p{#1pt}}
\newcolumntype{y}[1]{>{\raggedright\arraybackslash}p{#1pt}}
\newcolumntype{z}[1]{>{\raggedleft\arraybackslash}p{#1pt}}

\definecolor{grey}{rgb}{0.8, 0.8, 0.8}

\newcommand{\ccol}{\cellcolor{grey}}

\begin{document}
\maketitle

\begin{abstract}
Cross-modal matching, a fundamental task in bridging vision and language, has recently garnered substantial research interest. Despite the development of numerous methods aimed at quantifying the semantic relatedness between image-text pairs, these methods often fall short of achieving both outstanding performance and high efficiency. In this paper, we propose the crOss-Modal sInkhorn maTching (OMIT) network as an effective solution to effectively improving performance while maintaining efficiency. Rooted in the theoretical foundations of Optimal Transport, OMIT harnesses the capabilities of Cross-modal Mover's Distance to precisely compute the similarity between fine-grained visual and textual fragments, utilizing Sinkhorn iterations for efficient approximation. To further alleviate the issue of redundant alignments, we seamlessly integrate partial matching into OMIT, leveraging local-to-global similarities to eliminate the interference of irrelevant fragments. We conduct extensive evaluations of OMIT on two benchmark image-text retrieval datasets, namely  Flickr30K and MS-COCO. The superior performance achieved by OMIT on both datasets unequivocally demonstrates its effectiveness in cross-modal matching. Furthermore, through comprehensive visualization analysis, we elucidate OMIT's inherent tendency towards focal matching, thereby shedding light on its efficacy. Our code is publicly available at~\url{https://github.com/ppanzx/OMIT}.

\end{abstract}

\section{Introduction}
\label{sec:1-intro}

\begin{figure}[tbp]
    \centering
    \includegraphics[width=1\linewidth]{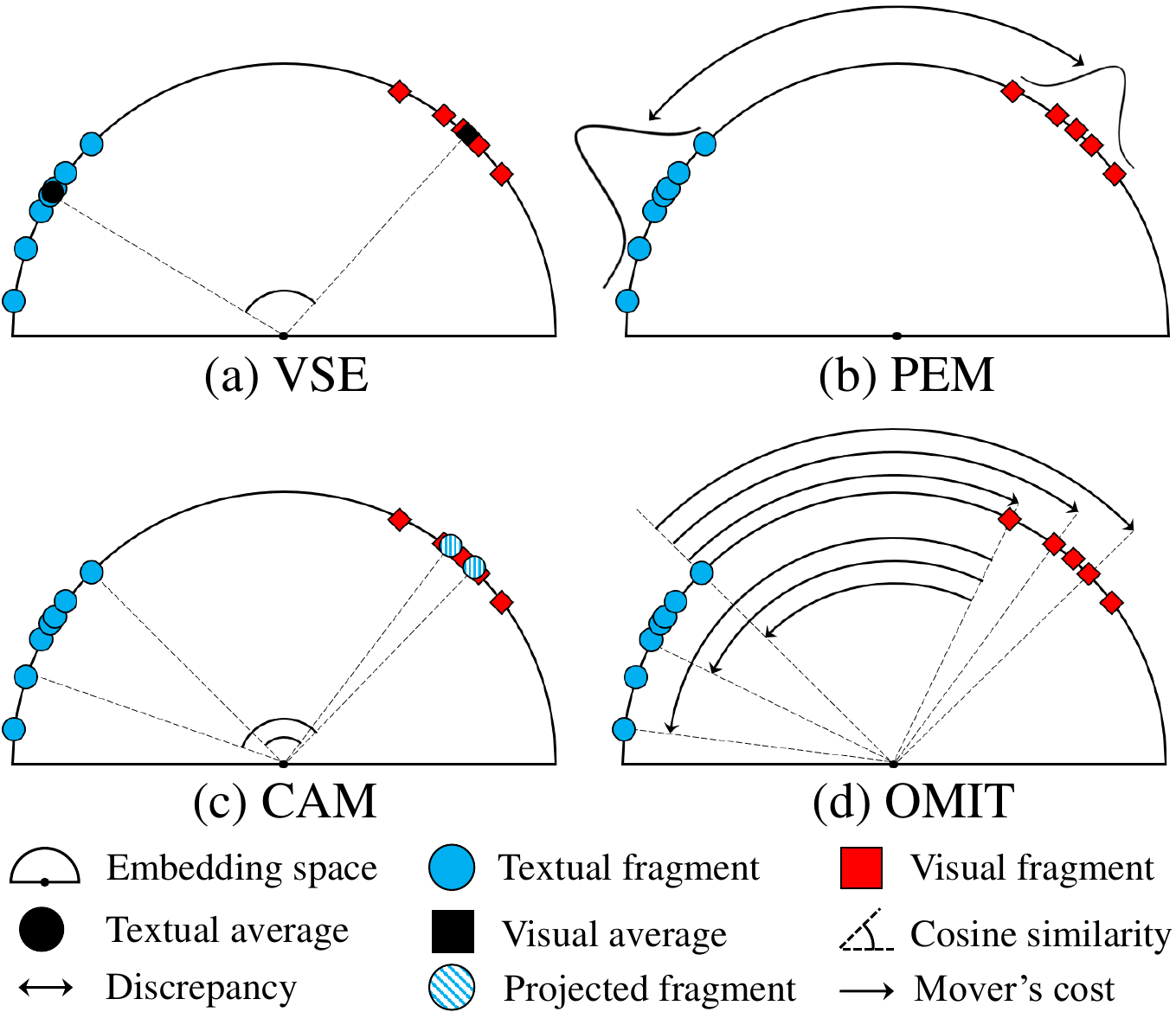}
    \caption{Comparison between different image-text matching methods. (a) VSE directly calculates the cosine similarity between two average vectors. (b) PEM substitutes the cosine similarity with the divergence of parametric distributions. (c) CAM measures the average cosine similarities between query fragments and their projections. (d) OMIT regards the Cross-modal Mover's Distance as the dissimilarity.}
    \label{fig:methods}
\end{figure}

This study delves into the essential task of image-text matching, at the intersection of vision and language, with the goal of establishing semantic connections between visual and textual data. The conventional method for image-text matching, known as Visual Semantic Embedding~\cite[VSE]{frome2013devise}, involves embedding data from different modalities into a single vector in a common representation space. This arrangement ensures that semantically coherent samples are positioned adjacently, as depicted in Figure~\ref{fig:methods}~(a). Despite VSE's notable improvements, recent research has underscored the insufficiency of representing images or captions with singular vectors due to inherent ambiguity~\cite{song2019polysemous}. Consequently, researchers have shifted their focus toward a more fine-grained form of cross-modal matching.

Existing fine-grained matching methods can be broadly categorized into two groups: the cross-attention-based method~(CAM) such as \cite{lee2018stacked}, and the probabilistic-embedding-based method~(PEM) like \cite{chun2021probabilistic}. PEM assumes that each sample corresponds to a Gaussian distribution within the common space. By incorporating an additional covariance, it extends the semantic information of the samples, leading to more comprehensive representations, as illustrated in Figure~\ref{fig:methods}~(b). The overall similarity between an image-text pair is then defined by measuring the distance between these two distributions. However, relying solely on the distribution assumption may produce degenerate results due to the presence of inter-dependency among local fragments and the lack of semantic continuity.

CAM assumes that an image can adequately depict a caption when the image-caption pair is semantically aligned, it also gauges the probability of textual fragment absence in the image through reconstruction error. Specifically, CAM projects each sample into a diverse feature set and computes the average projection similarity between the two feature sets, as depicted in Figure~\ref{fig:methods}(c). CAM effectively discovers shared semantics, thus presents improved performance, however, as discussed in~\cite{pan2023fine}, it is susceptible to redundant alignments and low efficiency. In addition, CAM fails to consider the relative significance among textual fragments, leading to limited performance.

Taking into account the simultaneous importance of both visual and textual fragments, we extend the concept of cross-attention to Optimal Transport (OT). As depicted in Figure 1(d), OT endeavors to minimize the cumulative cost of "transporting" embeddings from one modality to their corresponding embeddings in the other modality, describing a good metric for measuring the distance between two discrete distributions. The bidirectional nature of OT is reflected in the resulting transport plan, which accounts for the constraints of visual and textual significance. To address the issue of low computational efficiency in OT, we leverage the Sinkhorn-Knopp algorithms~\cite{cuturi2013sinkhorn} to approximate Earth Mover's Distance with the Sinkhorn distance, enabling both effective and efficient retrieval. Our experiments show that Sinkhorn distance serves as a more appropriate metric for capturing semantic correlations than current counterparts.

Additionally, to address the problem of redundant alignments, we introduce the concept of partial matching into OT, resulting in Optimal Partial Transport (OPT). In comparison to OT, OPT discards the requirement for complete correspondence and relaxes the marginal constraints, thereby enhancing its resilience to the challenges posed by noisy fragments. Specifically, we leverage the global embeddings as "dustbins" for matching, enhancing the alignments with common semantics and eliminating irrelevant alignments. Building on OPT, we introduce the crOss-Modal sInkhorn maTching~(OMIT) network. Remarkably, OMIT demonstrates superior capabilities in uncovering latent visual-semantic alignments, surpassing the performance of current state-of-the-art (SOTA) methods on two cross-modal retrieval benchmarks. In summary, our contributions can be succinctly outlined as follows:

(1) We address measuring the similarity between cross-modal pairs by harnessing the power of Optimal Transport. We demonstrate that the Sinkhorn distance, a numerical approximation of the Earth Mover's Distance, efficiently and effectively captures semantic correlations between visual and textual grained embeddings.

(2) We introduce OMIT for cross-modal retrieval, using cross-modal local-to-global correlations as a criterion to effectively filter out redundant alignments. Our OMIT achieves SOTA results on two benchmark datasets, i.e., MS-COCO~\cite{lin2014microsoft} and Flickr30K~\cite{young2014image}, demonstrating its effectiveness of partial matching.

(3) Our experiments shed light on the optimal transport plan generated by OMIT, showcasing its more focused attention compared to the counterpart generated by CAM. These findings align with existing literature~\cite{zhang2022negative}, further validating the efficacy of our approach.
\section{Related Works}
\label{sec:2-related}

\begin{figure*}
    \centering
    \includegraphics[width=1\linewidth]{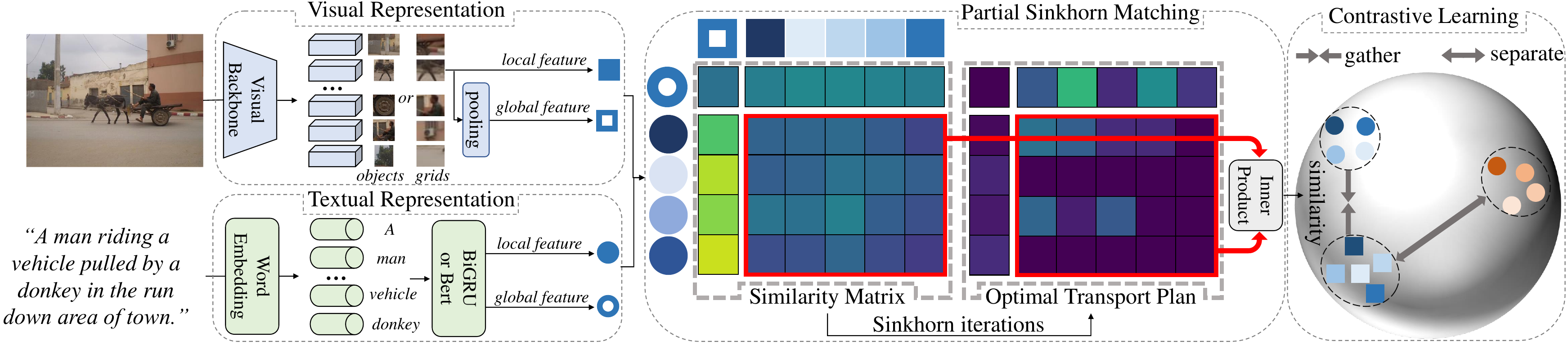}
    \caption{Overview of the OMIT. We begin by projecting the caption and image to the common space using textual and visual encoders, resulting in cross-modal fragments. Next, we employ the partial matching and calculate the Sinkhorn distance between fragments in the Partial Sinkhorn Matching module. Finally, we leverage the contrastive loss to group relevant fragments and disti nguish irrelevant fragments.}
    \label{fig:framework}
\end{figure*}

\mypara{Cross-modal Matching.} \cmmnt{Cross-modal matching presents a formidable challenge involving the alignment of corresponding concepts across different modalities. The intricacy arises from the necessity of precisely gauging the relevance of diverse representations, tantamount to sieving out grains of incongruent semantics. Given the labor-intensive nature of manually tagging all matching fragments, particularly for expansive datasets, reliance on approaches capable of inferring latent correspondences between modalities becomes imperative.

}An initial milestone in cross-modal matching was set by na\"ive VSE. This method hinges on merging local embeddings into a global embedding via various aggregation techniques. Examples include plain mean pooling~\cite{frome2013devise,faghri2017vse++}, self-attention layers~\cite{wang2020consensus}, GCN~\cite{li2019visual}, or learnable pooling~\cite{chen2021learning}. Subsequently, pairwise global embedding similarity is calculated using cosine similarity for ranking. However, VSE exhibits shortcomings in reasoning implicit associations and managing ambiguity. To address these limitations, PEM enhances VSE by employing probabilistic representations. It establishes a richer embedding space, where each image or text is represented as a Gaussian distribution, and similarity is computed as the divergence between these distributions. Subsequent research endeavors have further bolstered the efficacy of PEM through advancements in geometric spaces~\cite{wang2023multilateral}, multifaceted relationships~\cite{wang2023multilateral}, and superior metrics~\cite{an2023maximum}. Despite these efforts, the enhancements stemming from probabilistic embedding are modest, as evidenced by findings in~\cite{chun2021probabilistic}. Consequently, the representation of data with parametric distributions inherently exhibits limitations.

A distinctive approach, CAM, aligns discrete embeddings by leveraging set algebra guidance. It gauges whether a data sample (e.g., image) embodies the concept of a fragment from the other modality (e.g., token) by assessing the reconstruction similarity between the textual embedding and visual feature sets. Nevertheless, recent studies~\cite{pan2023fine} have illuminated that CAM fundamentally optimizes the coding distance and grapples with redundant alignment issues. In response, strategies to address redundancy have emerged, such as\cmmnt{ PFAN~\cite{liu2019focus}, which proposed focal attention based on inter-modality relations to eliminate irrelevant fragments from the shared semantic,} IMRAM~\cite{chen2020imram}, which iteratively refines coarse correspondences using cross-attention layers,\cmmnt{ SGRAF~\cite{diao2021similarity}, which uses GNN and self-attention to reason about meaningful relations,} NAAF~\cite{zhang2022negative}, which introduces a negative attention module to amplify the significance of matched fragments, and CHAN~\cite{pan2023fine}, which deconstructs the cross-attention mechanism with a coding framework to mitigate surplus alignments. Despite these advances, CAM's asymmetry and sub-optimality stem from its limited robustness in facilitating complete mutual expression. To surmount these challenges, our proposed OMIT adeptly sidesteps the issues elucidated above.

\mypara{Optimal Transport.} The lineage of research on Optimal Transport (OT) traces back to the 18th-century Monge problem, which aimed to transfer a heap of sand to a target pile of specific shape with minimal effort~\cite{peyre2019computational}. The minimal cost, also known as the Earth Mover's Distance (EMD) or Wasserstein distance, has found extensive applications in Computer Vision and Natural Language Processing for addressing matching challenges. For instance, \cite{rubner2000earth} utilized EMD to match the signatures of two images for image retrieval, \cite{kusner2015word} calculated the EMD between the embedded words of two text documents to indicate their dissimilarity. \cite{arjovsky2017wasserstein} minimized the EMD between real data distribution and a generator's distribution.\cmmnt{Sarlin et al.~\cite{sarlin2019coarse} introduced a matching network with the core of OT to find the correspondences between local features of pairwise images. \cite{flamary2016optimal} used EMD to perform the alignment of representations of source domain and target domain to resolve domain adaptation problems. Recent works have been motivated by~\cite{arjovsky2017wasserstein} and adopted Wasserstein distance to measure the discrepancy between distributions.}

The endeavor to extend OT from single-modality to cross-modality is a burgeoning field that has recently captivated researchers. \cite{chen2020graph} approached cross-modal matching problems from the perspective of graph matching, utilizing OT to simultaneously match nodes and edges between two heterogeneous graphs. \cite{wang2021wasserstein} aligned the graph of raw data and the counterpart graph in a graph dictionary with OT, thus achieving powerful feature embedding. \cite{zou2022tokenflow} deviated from the graph matching formulation and introduced a simple method to align image patches with text tokens using empirical approximate EMD. This innovation inspired our goal to directly aligning cross-modal fragments. Distinguishing itself from the approach in~\cite{zou2022tokenflow}, our work leverages the precision of the Bregman algorithm~\cite{bregman1967relaxation} for iterative EMD approximation, avoiding sub-optimal solution. 


While several works have utilized OT for cross-modal matching, they often focus solely on the full correspondence between local fragments, disregarding the fact that many alignments lack meaningful associations. To the best of our knowledge, we are the first to address this limitation by relaxing the constraint of full correspondence and enhancing the capabilities of OT through partial matching. 
\section{Approach}
\label{sec:3-method}

\subsection{Image-text Matching with Optimal Transport} 
Without loss of generality, an image can be represented granularly by a set of local embeddings denoted as $\mat{\calV}_m = \{\vct{v}_i|_{i=1}^{K_m}\}$, while a caption is represented as $\mat{\calT}_n = \{\vt_j|_{j=1}^{L_n}\}$. In this context, $m$ and $n$ index the dataset, $\vct{v}_i \in \R^d$ and $\vt_j \in \R^d$ represent the $d$-dimensional embeddings of an image sub-region/patch and a text token, respectively. Furthermore, $K_m$ and $L_n$ indicate the number of fragments in the image and the caption, respectively. Optimal Transport (OT) computes the Cross-modal Mover's Distance, denoted as $\mathcal{D}_{OT}$, between an image $\mathcal{V}_m$ and a caption $\mathcal{T}_n$ as a measure of their semantic similarity. The formulation of this distance is as follows:
\begin{align}
    \begin{split}
        \mathcal{D}_{OT}(\mat{\calV}_m, \mat{\calT}_n) = \mathop{min}\limits_{\mOmega\in\mU(\valpha,\vbeta)} \left\langle \mOmega,\mC \right\rangle\coloneqq\sum_{i=1}^{K_m}\sum_{j=1}^{L_n}\omega_{ij}c_{ij} \\
        \text{subject to}~~\mOmega\ones_{L_n} = \valpha,~\mOmega^\top\ones_{K_m} = \vbeta,~\mOmega\in \R_+^{K_m\times L_n} \\
    \end{split}
    \label{eq:dot}
    \centering
\end{align}

where $\valpha$ and $\vbeta$ denote marginal distributions of visual fragments and textual fragments, respectively. For simplicity, we treat all fragments within the same modality equally, thus setting $\valpha_i=\frac{1}{K_{m}}$ and $\vbeta_j=\frac{1}{L_{n}}$. $\mC\in\R_+^{K_m\times L_n}$ is the cost matrix, where each element $c_{ij}$ represents the distance between $\vct{v}_i$ and $\vt_j$. In this case, we choose $c_{ij}$ to be twice the $l_2$ distance: $c_{ij}=1-\vct{v}_i^\top \vt_j$. The matrix $\mOmega$ corresponds to one of the transport plans denoted as $\mU(\valpha, \vbeta)$, which must satisfy the marginal conditions in Equation~\ref{eq:dot}. $\omega_{ij}$ refers to an individual element of $\mOmega$. 

However, directly optimizing Equation~\ref{eq:dot} can be computationally intensive. To overcome this challenge, we adopt an approach proposed in~\cite{cuturi2013sinkhorn}, which introduces an entropic constraint on $\mOmega$ and constrains the feasible solution space, resulting in efficient optimization. As a result, we approximate the CMD by utilizing the Sinkhorn distance $\mathcal{D}_{S}$, which is given by:
\begin{align}
    \centering
        \mathcal{D}_{S}(\mat{\calV}_m, \mat{\calT}_m) &= \mathop{min}\limits_{\mOmega\in\mU(\valpha,\vbeta)} \left\langle \mOmega,\mC \right\rangle -\lambda H(\mOmega)
    \label{eq:dsink}
\end{align}
where $H(\mOmega)=-\sum_{i=1}^{K_m}\sum_{j=1}^{L_n}\omega_{ij}(\log \omega_{ij}-1)$ is the entropy term, and $\lambda\geq 0$ is the scalar factor that balances the convergence speed and proximity to the optimal solution. Mathematically, when $\lambda$ is sufficiently small, the solution to Equation~\ref{eq:dsink} is equivalent to the optimal plan derived from Equation~\ref{eq:dot}. The unique solution $\mOmega^*$ of Equation~\ref{eq:dsink} can be obtained by iteratively projecting the Gibbs kernel $\mxi=e^{-\frac{\mC}{\lambda}}$ to normalize either its rows or columns using Sinkhorn iterations~\cite{cuturi2013sinkhorn,benamou2015iterative}:
\begin{equation}
    \centering
    \begin{split}
        &\mOmega^{(2t+1)} = \mathop{diag}(\frac{\valpha}{\mOmega^{(2t)}\ones_{L_n}})\mOmega^{(2t)} \\
        \mOmega^{(2t+2)} &= \mOmega^{(2t+1)}\mathop{diag}(\frac{\vbeta}{\mOmega^{(2t+1)\top}\ones_{K_m}}) \\
        &\text{Initialize}~~\mOmega^{(0)}=\mxi
    \end{split}
    \label{eq:projection}
\end{equation}
where $t$ denotes the iteration steps. Equation~\ref{eq:projection} will converge with several steps, ensuring efficient matching. 

\subsection{Improving OT with Partial Matching}
One evident challenge in vanilla OT for image-text matching is the presence of redundant alignments. These alignments refer to region-word pairs that lack semantic coherence but contribute significantly to the overall similarity. Since an image and a sentence cannot fully capture each other's content~\cite{yang2022vision}, there are instances where certain objects in the image do not correspond to any words in the sentence. When employing na\"ive OT for matching, these pairs are inevitably included in the overall similarity computation, resulting in unwanted disturbances. A straightforward solution to this problem is to establish a suitable threshold for each image-text pair, filtering out alignments with similarities below the threshold. However, determining the appropriate threshold for each fragment requires fragment-specific considerations, making manual tuning a time-consuming and labor-intensive process.

Inspired by the concept of partial matching in Optimal Partial Transport, we have discovered that re-weighting is a more appropriate and efficient approach compared to thresholding. Instead of filtering out redundant alignments, we assign them a small weight, thereby reducing their interference in the matching process. To achieve this, we introduce additional visual global embedding $\overline{\vct{v}}$ and textual embedding $\overline{\vt}$ as "dustbins" for matching. In this manner, we redefine the optimal transport plan as follows:
\begin{align}
    \begin{split}
    &\hat{\mOmega}^* = \mathop{arg~min}\limits_{\hat{\mOmega}\in\mU(\hat{\valpha},\hat{\vbeta})} \left\langle \hat{\mOmega},\hat{\mC} \right\rangle \\
    &= \mathop{arg~min}\limits_{\hat{\mOmega}\in\mU(\hat{\valpha},\hat{\vbeta})} \left\langle \left[
         \begin{array}{cc}
             \mOmega_{g} & \mOmega_{gv} \\
             \mOmega_{gt} & \mOmega 
         \end{array}
     \right],
     \left[
         \begin{array}{cc}
             \mC_{g} & \mC_{gv} \\
             \mC_{gt} & \mC
         \end{array}
     \right] \right\rangle \\
        &\text{subject to}~~\hat{\mOmega}\ones_{L_n+1} = \hat{\valpha},~\hat{\mOmega}^\top\ones_{K_m+1} = \hat{\vbeta},\\
        &~~~~~~~~~~~~~~~~\hat{\mOmega}\in \R_+^{(K_m+1)\times (L_n+1)} \\
    \end{split}
    \label{eq:opt}
    \centering
\end{align}
The symbols superscript with $\hat{}$ denote the extended version of their original matrix. For instance, $\hat{\mC}$ extends $\mC$ by incorporating auxiliary components such as the cost between dustbins $\mC_{g}$, the cost between visual dustbin and textual fragments $\mC_{gv}$ and the cost between textual dustbin and visual fragments $\mC_{gt}^*$. Equation~\ref{eq:opt} can be solved using Sinkhorn iterations formulated in Equation~\ref{eq:projection}. The overall similarity between the image $\mat{\calV}_m$ and the caption $\mat{\calT}_m$ is defined as:
\begin{equation}
    \mathcal{S}_{OT}(\mat{\calV}_m, \mat{\calT}_m) = 1-\mathcal{D}_{OT} = \left\langle \mOmega^*,\mat{\calV}_m\mat{\calT}_n^{\top} \right\rangle
    \label{eq:sot}
\end{equation}

In above equation, we consider only the optimal transport plan between local fragments to quantify the overall similarity. We discard the auxiliary parts such as $\mOmega_{g}^*$, $\mOmega_{gv}^*$, and $\mOmega_{gt}^*$. By doing so, we ensure that fragments irrelevant to the shared semantics correspond to the global dustbin, effectively eliminating their contribution to the overall similarity.
\begin{table*}[t]
    \centering
    \small
    \caption{Image-Text Retrieval Results of OMIT method on Flickr 30K and COCO 5K test set, using different visual and text backbones (denoted by \textbf{bold section title}). $\star$: Ensemble results of two models. $\dag$ denotes results using post-processing techniques like DSL~\cite{cheng2021improving}. The best (in \symtext{rsum}) are marked \textbf{bold} in \first{red}.}
    \label{tab:f30k&coco_full}
    {
        \tabcolsep 3 pt
	    \begin{tabular}{@{\;} l c ccccccc ccccccc@{\;}}
	        \addlinespace
	        \toprule
	        \multicolumn{1}{c}{\multirow{3}{*}{\symtext{Method}}} & \multicolumn{1}{c}{\multirow{3}{*}{\symtext{Type}}} & \multicolumn{7}{c}{ Flickr 30K Test set~\cite{young2014image}} &
	        \multicolumn{7}{c}{COCO 5$\mathrm{K}$Test set~\cite{lin2014microsoft}} \\ \cmidrule(lr){3-9} \cmidrule(lr){10-16}
	        {} & {} & \multicolumn{3}{c}{\textsc{img}~$\rightarrow$~\textsc{text}} & \multicolumn{3}{c}{\textsc{text}~$\rightarrow$~\textsc{img}} & \multicolumn{1}{c}{\multirow{2}{*}{\symtext{rsum}~$\uparrow$}}& \multicolumn{3}{c}{\textsc{img}~$\rightarrow$~\textsc{text}} & \multicolumn{3}{c}{\textsc{text}~$\rightarrow$~\textsc{img}} & \multicolumn{1}{c}{\multirow{2}{*}{\symtext{rsum}~$\uparrow$}} \\ 
	        \cmidrule(lr){3-5} \cmidrule(lr){6-8} \cmidrule(lr){10-12} \cmidrule(lr){13-15}
	        {} & {} & \symtext{r}{@1} & \symtext{r}{@5} & \symtext{r}{@10} & \symtext{r}{@1} & \symtext{r}{@5} & \symtext{r}{@10} &  & \symtext{r}{@1} & \symtext{r}{@5} & \symtext{r}{@10} & \symtext{r}{@1} & \symtext{r}{@5} & \symtext{r}{@10} &   \\
            \midrule
			\multicolumn{9}{@{\;}l}{\bf {ResNet-152~\cite{he2016deep} $+$ BiGRU~\cite{chung2014empirical}}} \\[1.5pt]
    		VSE++~\cite{faghri2017vse++} & VSE & 52.9 & 80.5 & 87.2 & 39.6 & 70.1 & 79.5 & 409.8 & 41.3 & 71.1 & 81.2 & 30.3 & 59.4 & 72.4 & 355.7 \\
                PVSE~\cite{song2019polysemous} & PEM & 59.1 & 84.5 & 91.0 & 43.4 & 73.1 & 81.5 & 432.6 & 45.2 & 74.3 & 84.5 & 32.4 & 63.0 & 75.0 & 374.4 \\
                PCME~\cite{chun2021probabilistic} & PEM & 58.5 & 81.4 & 89.3 & 44.3 & 72.7 & 81.9 & 428.1 & 44.2 & - & - & 31.9 & - & - & - \\
                SBE~\cite{kim2023improving} & CAM & 61.8 & 85.5 & 91.1 & 46.1 & 74.8 & 83.3 & 442.6 & 47.2 & 74.8 & 84.1 & 33.8 & 63.1 & 74.7 & 377.7 \\
                \rowcolor{Gray} ours & OMIT & 67.6 & 90.6 & 95.2 & 52.5 & 79.3 & 86.6 & \first{471.8} & 46.7 & 76.9 & 87.0 & 35.9 & 65.4 & 76.7 & \first{388.6} \\
			\midrule
	        \multicolumn{9}{@{\;}l}{\bf {BUTD~\cite{anderson2018bottom} $+$ BiGRU~\cite{chung2014empirical}}} \\[1.5pt]
                 SCAN$\star$~\cite{lee2018stacked} & CAM & 67.4 & 90.3 & 95.8 & 48.6 & 77.7 & 85.2 & 465.0 & 50.4 & 82.2 & 90.0 & 38.6 & 69.3 & 80.4 & {410.9} \\
	           VSRN$\star$~\cite{li2019visual} & VSE & 71.3 & 90.6 & 96.0 & 54.7 & 81.8 & 88.2 & 482.6 & 53.0 & 81.1 & 89.4 & 40.5 & 70.6 & 81.1 & {415.7}  \\
            CAAN~\cite{zhang2020context} & VSE & 70.1 & 91.6 & 97.2 & 52.8 & 79.0 & 87.9 & 478.6 & 52.5 & 83.3 & 90.9 & 41.2 & 70.3 & 82.9 & {421.1} \\
       	    IMRAM$\star$~\cite{chen2020imram} & CAM & 74.1 & 93.0 & 96.6 & 53.9 & 79.4 & 87.2 & {484.2} & 53.7 & 83.2 & 91.0 & 39.7 & 69.1 & 79.8 & {416.5} \\
                SGRAF$\star$~\cite{diao2021similarity} & CAM & 78.4 & 94.6 & 97.5 & 58.2 & 83.0 & 89.1 & {500.8} & 55.8 & 83.0 & 91.0 & 42.0 & 72.4 & 82.1 & 426.3 \\
                VSE$\infty$~\cite{chen2021learning} & VSE & 76.5 & 94.2 & 97.7 & 56.4 & 83.4 & 89.9 & 498.1 & 56.6 & 83.6 & 91.4 & 39.3 & 69.9 & 81.1 & {421.9} \\
                NAAF~\cite{zhang2022negative} & CAM & 79.6 & 96.3 & 98.3 & 59.3 & 83.9 & 90.2 & {507.6} & 58.9 & 85.2 & 92.0 & 42.5 & 70.9 & 81.4 & {430.9} \\
                SBE~\cite{kim2023improving} & PEM & 77.8 & 94.0 & 97.5 & 57.5 & 84.0 & 90.0 & 500.8 & 58.8 & 84.9 & 91.5 & 41.1 & 72.0 & 82.4 & 430.7 \\
                HREM~\cite{fu2023learning} & VSE & 79.5 & 94.3 & 97.4 & 59.3 & 85.1 & 91.2 & 506.8 & 58.9 & 85.3 & 92.1 & 40.0 & 70.6 & 81.2 & 428.1 \\
             \rowcolor{Gray} ours & OMIT & 80.1 & 96.1 & 98.2 & 61.9 & 85.8 & 91.5 & \first{513.5} & 60.2 & 85.6 & 92.1 & 42.6 & 71.8 & 81.4 & \first{433.8} \\
	        \midrule
	        \multicolumn{9}{@{\;}l}{\bf {BUTD~\cite{anderson2018bottom} $+$ {BERT}~\cite{devlin2018bert}}} \\[1.5pt]
	        MMCA~\cite{wei2020multi} & CAM & 74.2 & 92.8 & 96.4 & 54.8 & 81.4 & 87.8 & {487.4} & 54.0 & 82.5 & 90.7 & 38.7 & 69.7 & 80.8 & {416.4}\\
	        VSE$\infty$~\cite{chen2021learning} & VSE & 81.7 & 95.4 & 97.6 & 61.4 & 85.9 & 91.5 & {513.5} & 58.3 & 85.3 & 92.3 & 42.4 & 72.7 & 83.2 & {434.3} \\
            TERAN$\star$~\cite{messina2021fine} & CAM & 79.2 & 94.4 & 96.8 & 63.1 & 87.3 & 92.6 & {513.4} & 59.3 & 85.8 & 92.4 & 45.1 & 76.4 & 84.4 & {443.4} \\
            VSRN++$\star$~\cite{li2022image} & VSE & 79.2 & 94.6 & 97.5 & 60.6 & 85.6 & 91.4 & {508.9} & 54.7 & 82.9 & 90.9 & 42.0 & 72.2 & 82.7 & {425.4} \\
	      CHAN~\cite{pan2023fine}& CAM & {80.6} & {96.1} & {97.8} & {63.9} & {87.5} & {92.6} & {518.5} & {59.8} & {87.2} & {93.3} & {44.9} & {74.5} & {84.2} & {443.9} \\
            HREM~\cite{fu2023learning} & VSE & 83.3 & 96.0 & 98.1 & 63.5 & 87.1 & 92.4 & 520.4 & 62.3 & 87.6 & 93.4 & 43.9 & 73.6 & 83.3 & 444.1 \\
            \rowcolor{Gray} ours & OMIT & 83.1 & 96.5 & 98.7 & 65.7 & 88.5 & 93.3 & \first{525.8} & 64.5 & 88.2 & 93.7 & 46.0 & 75.3 & 84.4 & \first{452.0} \\
            \midrule
            \multicolumn{9}{@{\;}l}{\bf {ResNeXT-101(WSL)~\cite{mahajan2018exploring} $+$ {BERT}~\cite{devlin2018bert}}} \\[1.5pt]
            VSE$\infty$~\cite{chen2021learning} & VSE & 88.4 & 98.3 & 99.5 & 74.2 & 93.7 & 96.8 & 550.9 & 66.4 & 89.3 & 94.6 & 51.6 & 79.3 & 87.6 & 468.9 \\
            SBE~\cite{kim2023improving} & PEM & 88.8 & 98.5 & 99.6 & 74.3 & 94.0 & 96.7 & 551.9 & 69.1 & 90.7 & 95.6 & 52.1 & 79.6 & 87.8 & 474.9 \\
	      \rowcolor{Gray} ours & OMIT & 91.0 & 99.0 & 99.9 & 78.1 & 95.2 & 97.6 & \first{560.7} & 69.0 & 90.7 & 95.5 & 54.0 & 80.4 & 88.3 & \first{477.9}\\
            \midrule
            \multicolumn{9}{@{\;}l}{\bf {CLIP VIT-B/32~\cite{radford2021learning} }} \\[1.5pt]
            zero-shot CLIP~\cite{radford2021learning} & VSE & 78.9 & 94.9 & 98.2 & 58.8 & 83.5 & 90.0 & 504.3 & 50.1 & 74.9 & 83.4 & 30.4 & 56.0 & 66.9 & 361.7 \\
            fine-tuned CLIP & VSE & 86.9 & 97.3 & 99.3 & 72.0 & 92.3 & 96.2 & 544.1 & 62.9 & 85.8 & 92.0 & 46.2 & 73.4 & 82.9 & 443.3 \\
            \rowcolor{Gray} ours & OMIT & 86.4 & 97.8 & 99.7 & 73.6 & 92.8 & 96.0 & 546.2 & 63.3 & 86.7 & 92.5 & 48.5 & 75.7 & 84.2 & 450.9 \\
            \rowcolor{Gray} ours$\dag$& OMIT & 93.9 & 99.1 & 99.6 & 76.6 & 93.9 & 97.1  & \first{560.2} & 72.1 & 89.3 & 93.9  & 50.1 & 75.9 & 84.3 & \first{465.6} \\
            
            \bottomrule
	    \end{tabular}
	}

\end{table*}

\subsection{Cross-Modal Sinkhorn Matching}
The OMIT framework comprises four components, as depicted  in Figure~\ref{fig:framework}. The Partial Sinkhorn Matching module has been primarily discussed earlier, and the remaining modules are described in detail below.

\mypara{Visual Representation.} We adopt two different types of visual features, namely grid feature~\cite{jiang2020defense} and region feature~\cite{anderson2018bottom}, for granular representation. For grid representation, we employ a CNN like ResNet-152~\cite{he2016deep} or ResNeXT-101~\cite{xie2017aggregated} as the backbone (excluding the last pooling and linear layer) to capture image patches. Alternatively, to capture higher-level information such as objects within an image, we utilize the Faster-RCNN~\cite{ren2017faster} equipped with Bottom-Up-and-Top-Down attention (BUTD)~\cite{anderson2018bottom}, which is pretrained on the Visual Genome dataset~\cite{krishna2017visual}, to encode salient sub-regions. Following nonlinear transformations and normalization, the resulting grid or region features are considered as the visual fragments $\{\vct{v}_i|_{i=1}^{K_m}\}$. To obtain the global feature $\overline{\vct{v}}$, we simply apply average pooling to the fragments.

\mypara{Text Representation.} To obtain the textual fragments $\{\vt_j|_{j=1}^{L_n}\}$, we employ sequential networks such as BiGRU~\cite{chung2014empirical} or pretrained Bert~\cite{devlin2018bert} to contextualize dense word embeddings~\cite{pennington2014glove}. Each output from the sequential model represents a granular embedding~$\vt_j$. Subsequently, we project $\vt_j$ into the same space as $\vct{v}_i$ to enable subsequent matching. When using BERT, we adopt the \textbf{CLS} token as the global feature. For BiGRU, we compute the mean vector of the local embeddings to obtain the global feature.

\mypara{Objective.} Following previous works~\cite{faghri2017vse++,lee2018stacked,li2019visual}, we adopt a hinge-based triplet ranking loss with online hardest negative mining~\cite{faghri2017vse++} to enforce that the similarity between a semantically consistent image-text pair is greater than that of an unrelated image-text pair by a margin $\phi$. The optimization objective is defined by the following loss function:
\begin{align}
    \label{eq:objective}
    &\calL = \sum_{m=1}^{M}\mathop{max}\limits_{n\neq m}~[\phi + \mathcal{S}_{OT}(\mat{\calV}_m, \mat{\calT}_m) - \mathcal{S}_{OT}(\mat{\calV}_m, \mat{\calT}_n)]_{+} \notag \\
            &+ \sum_{n=1}^{N} \mathop{max}\limits_{m\neq n}~[\phi + \mathcal{S}_{OT}(\mat{\calV}_n, \mat{\calT}_n) -\mathcal{S}_{OT}(\mat{\calV}_m, \mat{\calT}_n)]_{+}
\end{align}
where $M$ and $N$ are the sizes of images and captions in the minibatch. The notation $[~\cdot~]_{+}$ denotes the hinge function. Data with the same index signifies semantic consistency, whereas different indices indicate a mismatched pair.

\section{Experiments}
\label{sec:4-exper}

\begin{table}[ht]
    \centering
    \small
    \caption{Image-Text Retrieval Results of the OMIT on MS-COCO 5-fold 1K test set. }
    \label{tab:coco1k}
    {
        \tabcolsep 1.6 pt
	    \begin{tabular}{@{\;} l c ccccccc c@{\;}}
	        \addlinespace
	        \toprule
	        \multicolumn{1}{c}{\multirow{2}{*}{\symtext{Method}}}& \multicolumn{1}{c}{\multirow{2}{*}{\symtext{Type}}} &  \multicolumn{3}{c}{\textsc{img}~$\rightarrow$~\textsc{text}} &      \multicolumn{3}{c}{\textsc{text}~$\rightarrow$~\textsc{img}} & \multicolumn{1}{c}{\multirow{2}{*}{\symtext{rsum}~$\uparrow$}} & \\ 
	        \cmidrule(lr){3-5} \cmidrule(lr){6-8}
	        {} & {} & \symtext{r}{@1} & \symtext{r}{@5} & \symtext{r}{@10} & \symtext{r}{@1} & \symtext{r}{@5} & \symtext{r}{@10} &  \\
			\midrule
			\multicolumn{9}{@{\;}l}{\bf {ResNet-152~\cmmnt{he2016deep} $+$ BiGRU~\cmmnt{chung2014empirical}}} \\[3pt]
			VSE++\cmmnt{faghri2017vse++}& VSE & 64.6 & 90.0 & 95.7 & 52.0 & 84.3 & 92.0 & 478.6 \\
                PVSE\cmmnt{song2019polysemous} & PEM & 69.2 & 91.6 & 96.6 & 55.2 & 86.5 & 93.7 & 492.8 \\
                PCME\cmmnt{chun2021probabilistic} & PEM & 68.8 & - & - & 54.6 & - & - & - \\
                SBE\cmmnt{kim2023improving} & PEM & 70.3 & 91.5 & 96.3 & 56.0 & 85.8 & 93.3 & 493.2 \\
                \rowcolor{Gray} ours & OMIT & 71.3 & 93.6 & 97.7 & 57.8 & 87.2 & 94.2 & \first{501.9} \\
                
			\midrule
	        \multicolumn{8}{@{\;}l}{\bf {BUTD~\cmmnt{anderson2018bottom} $+$ BiGRU~\cmmnt{chung2014empirical}}} \\[3pt]
	        SCAN$\star$\cmmnt{lee2018stacked} & CAM & 72.7 & 94.8 & 98.4 & 58.8 & 88.4 & 94.8 & {507.9} \\
	        VSRN$\star$\cmmnt{li2019visual} & VSE & 76.2 & 94.8 & 98.2 & {62.8} & 89.7 & 95.1 & {516.8} \\
    	   CAAN~\cmmnt{zhang2020context} & VSE & 75.5 & 95.4 & 98.5 & 61.3 & 89.7 & 95.2 & {515.6} \\
      	    IMRAM$\star$\cmmnt{chen2020imram} & CAM & 76.7 & 95.6 & 98.5 & 61.7 & 89.1 & 95.0 & 516.6 \\
                SGRAF$\star$\cmmnt{diao2021similarity} & CAM & 79.6 & 96.2 & 98.5 & 63.2 & 90.7 & 96.1 & 524.3 \\
                VSE$\infty$\cmmnt{chen2021learning} & VSE & 78.5 & 96.0 & 98.7 & 61.7 & 90.3 & 95.6 & 520.8 \\
                NAAF~\cmmnt{zhang2022negative}  & CAM & 78.1 & 96.1 & 98.6 & 63.5 & 89.6 & 95.3 & {521.2} \\
                HREM~\cmmnt{fu2023learning} & VSE & 80.0 & 96.0 & 98.7 & 62.7 & 90.1 & 95.4 & 522.8 \\
                \rowcolor{Gray} ours & OMIT & 80.4 & 96.0 & 98.6 & 63.8 & 90.1 & 95.6 & \first{524.4} \\
	        \midrule
	        \multicolumn{8}{@{\;}l}{\bf {BUTD~\cmmnt{anderson2018bottom} $+$ {BERT} \cmmnt{devlin2018bert}}} \\[3pt]
	        MMCA \cmmnt{wei2020multi} & CAM & 74.8 & 95.6 & 97.7 & 61.6 & 89.8 & 95.2 & {514.7} \\
	        VSE$\infty$\cmmnt{chen2021learning} & VSE & 79.7 & 96.4 & 98.9 & 64.8 & 91.4 & 96.3 & {527.5} \\
                TERAN$\star$\cmmnt{messina2021fine} & CAM & 80.2 & 96.6 & 99.0 & 67.0 & 92.2 & 96.9 & {531.9} \\
                VSRN++$\star$~\cmmnt{li2022image} & VSE & 77.9 & 96.0 & 98.5 & 64.1 & 91.0 & 96.1 & {523.6} \\
	          CHAN~\cmmnt{pan2023fine} & CAM & {81.4} & {96.9} & {98.9} & {66.5} & {92.1} & {96.7} & 532.6 \\
                HREM~\cmmnt{fu2023learning} & VSE & 81.1 & 96.6 & 98.9 & 66.1 & 91.6 & 96.5 & 530.7 \\
                \rowcolor{Gray} ours & OMIT & 83.2 & 97.0 & 98.9 & 67.5 & 91.9 & 96.5 & \first{535.1} \\
                \midrule
                \multicolumn{9}{@{\;}l}{\bf {ResNeXT-101(WSL)~\cmmnt{mahajan2018exploring} $+$ {BERT} \cmmnt{devlin2018bert}}} \\[1.5pt]
                VSE$\infty$\cmmnt{chen2021learning} & VSE & 84.2 & 97.8 & 99.3 & 71.9 & 93.9 & 97.5 & 544.6 \\
                SBE\cmmnt{kim2023improving} & PEM & 86.3 & 97.8 & 99.4 & 72.4 & 94.0 & 97.6 & 547.5 \\
    	        \rowcolor{Gray} ours & OMIT & \ccol 85.7 & 98.1 & 99.3 & 73.2 & 94.4 & 97.8 & \first{548.5} \\
                \midrule
                \multicolumn{9}{@{\;}l}{\bf {CLIP VIT-B/32}} \\[1.5pt]
                zs CLIP & VSE & 69.1 & 91.0 & 95.7 & 49.7 & 79.4 & 88.8 & 473.6 \\
                ft CLIP & VSE & 80.8 & 96.2 & 98.2 & 66.3 & 91.0 & 96.4 & 528.8 \\
                \rowcolor{Gray} ours & OMIT & 81.6 & 96.7 & 98.6 & 68.2 & 91.7 & 96.2 & \first{532.9} \\
	       \bottomrule
	    \end{tabular}
	}
\end{table}

\subsection{Settings.} 

\mypara{Datasets.} We evaluate our approach on two benchmark image-text retrieval datasets, namely MS-COCO~\cite{lin2014microsoft} and Flickr30K~\cite{young2014image}, to validate its effectiveness. MS-COCO consists of 123,287 images, each of which is manually captioned with five sentences. Following the protocols in~\cite{faghri2017vse++}, we split 113,287 images in MS-COCO for training, 5,000 for validation, and the remaining 5,000 for testing. Similarly, images in Flickr30K are divided into 29,769 training images, 1,014 validating images, and 1,000 testing images.

\mypara{Metrics.} We employ the commonly used evaluation measure, $R@K$, which measures the percentage of relevant items among the top-K retrieved items for a given query, to assess the performance of our models. We consider $R@1$, $R@5$, and $R@10$ for universal evaluation and use $RSUM=R@1+R@5+R@10$ to compare overall performance between our method and others. All our results are reported on the test set using the model that performs best on the validation set. Notably, we do not report any integrated results by assembling two separated models but display the \textbf{single-model} results. 

\mypara{Implementation details.} The OMIT model is meticulously optimized using the AdamW optimizer~\cite{loshchilov2017decoupled} with a weight decay of $5e-4$. For the region feature, a batch size of 384 is utilized, while for the grid feature, a batch size of 128 is employed. where the initial 10 epochs adopt a learning rate of $5e-4$, followed by the subsequent 5 epochs with a learning rate of $5e-5$.. The resolution for the input image is set to 224$\times$224, the number of proposals for salient regions is set to 36. The values of $\lambda$ and $\phi$ are set to 0.02 and 0.05, respectively. $\tau$ in Equation~\ref{eq:opt} is set to 0.1. The visual mask ratio is set to 0.4 for the region feature and 0.2 for the grid feature. Captions are randomly masked at a ratio of 0.1, replaced at a ratio of 0.02, and discarded at a ratio of 0.08. Consistent with prior work~\cite{faghri2017vse++}, the embedding dimension $d$ is set to 1024. To address the issue of slow convergence in the ranking loss~\cite{wei2021universal}, we adopt the approach proposed in~\cite{chen2021learning} and perform a warm-up phase for OMIT during the first epoch, without engaging in hard negative mining.

\subsection{Comparison Results} 

Table~\ref{tab:f30k&coco_full} and Table~\ref{tab:coco1k} provide comprehensive comparisons of our OMIT with current state-of-the-art~(SOTA) methods on three benchmarks. For fairness, we categorize the methods according to their backbones. Remarkably, across all datasets and backbones, our OMIT outperforms all other methods, achieving the highest $RSUM$ score. Interestingly, all our text-to-image $R@1$ scores are approximately 2\% higher than those of current methods, likely because all our OMIT variants effectively harness the potential of visual backbones.

Specifically, when compared to baselines employing ResNet and BiGRU~(first blocks), OMIT exhibits significant improvements of at least 29.2\%, 10.9\% and 8.7\% over SOTA SBE in terms of $RSUM$ on the Flickr30K, COCO 5K and COCO 5-fold 1K, respectively. Furthermore, when compared to methods utilizing region features such as SCAN, NAAF, and HREM~(second and third blocks), OMIT achieves the best results at $R@1$ and $RSUM$ for both datasets under the same conditions of BUTD and BiGRU/Bert. Finally, when equipped with the more powerful visual encoder ResNeXT-101, along with a higher visual resolution~(last blocks), OMIT achieves significant retrieval improvements and consistently outperforms counterparts such as VSE$\infty$ and SBE.
\begin{table}[tbp]
    \centering
    \small
    \caption{Comparison of OMIT with other matching methods on the Flickr30K test set. The \symtext{time}~denotes the overall computational time for matching total visual embeddings and textual embeddings.}
    \label{tab:cmpr_mtch}
    {
        \tabcolsep 2 pt
	    \begin{tabular}{@{\;} l c cccccc c@{\;}}
	        \addlinespace
	        \toprule
	        \multicolumn{1}{c}{\multirow{2}{*}{\symtext{TYPE}}} &  \multicolumn{3}{c}{\textsc{img}~$\rightarrow$~\textsc{text}} &      \multicolumn{3}{c}{\textsc{text}~$\rightarrow$~\textsc{img}} & \multicolumn{1}{c}{\multirow{2}{*}{\symtext{rsum}$\uparrow$}} & \multicolumn{1}{c}{\multirow{2}{*}{\symtext{time}(s)$\downarrow$}}\\ 
	        \cmidrule(lr){2-4} \cmidrule(lr){5-7}
	        {} & \symtext{r}{@1} & \symtext{r}{@5} & \symtext{r}{@10} & \symtext{r}{@1} & \symtext{r}{@5} & \symtext{r}{@10} & \\
                \midrule
                \multicolumn{9}{@{\;}l}{\bf {ResNet-152 \cite{he2016deep} $+$ BiGRU~\cite{chung2014empirical}}} \\[1.5pt]
			VSE & 59.6 & 85.3 & 90.6 & 44.5 & 74.5 & 83.2 & 437.8 & 9.48\\
                PEM & 60.2 & 85.8 & 91.4 & 46.6 & 74.7 & 83.6 & 442.3 & \first{9.43} \\
                CAM & 67.4 & 91.1 & 95.3 & 50.5 & 78.5 & 86.8 & 469.7 & 42.55 \\
                \rowcolor{Gray} OMIT & 67.6 & 90.6 & 95.2 & 52.5 & 79.3 & 86.6 & \first{471.8} & 15.03 \\
                \midrule
	        \multicolumn{9}{@{\;}l}{\bf {BUTD~\cite{anderson2018bottom} $+$ BiGRU~\cite{chung2014empirical}}} \\[3pt]
         	VSE & 61.3 & 85.0 & 92.3 & 44.5 & 75.3 & 83.8 & 442.2 & \first{10.91} \\
                PEM & 70.7 & 90.8 & 94.8 & 51.2 & 79.4 & 86.7 & 473.6 & 12.03\\
                CAM & 80.5 & 95.3 & 97.7 & 58.7 & 84.2 & 90.3 & 506.8 & 22.31 \\
                \rowcolor{Gray} OMIT & 80.1 & 96.1 & 98.2 & 61.9 & 85.8 & 91.5 & \first{513.5} & 16.93 \\
	       \bottomrule
	    \end{tabular}
	}
\end{table}

\subsection{Ablation Study}
\mypara{Compared With Other Matching Methods.} We conduct experiments to perform a fair comparison between our OMIT and other existing methods. To ensure fairness, we exclusively replace the Partial Sinkhorn Matching module with various alternatives. The results are presented in Table~\ref{tab:cmpr_mtch}, where OMIT demonstrates the highest accuracy while remaining competitive in terms of efficiency. In particular, when compared to global embedding methods like VSE and PEM, OMIT exhibits noteworthy improvements in all effectiveness indicators, including $R@1$ and $RSUM$, despite a slight decrease in efficiency. In contrast, when pitted against other more precise methods like CAM, OMIT not only improves retrieval efficiency but also achieves superior accuracy. This enhanced efficiency of OMIT can be attributed to its omission of the reconstruction process, while the rationale behind its improved accuracy will be expounded upon in subsequent sections.

\begin{figure}[tbp]
    \centering
    \includegraphics[width=1\linewidth]{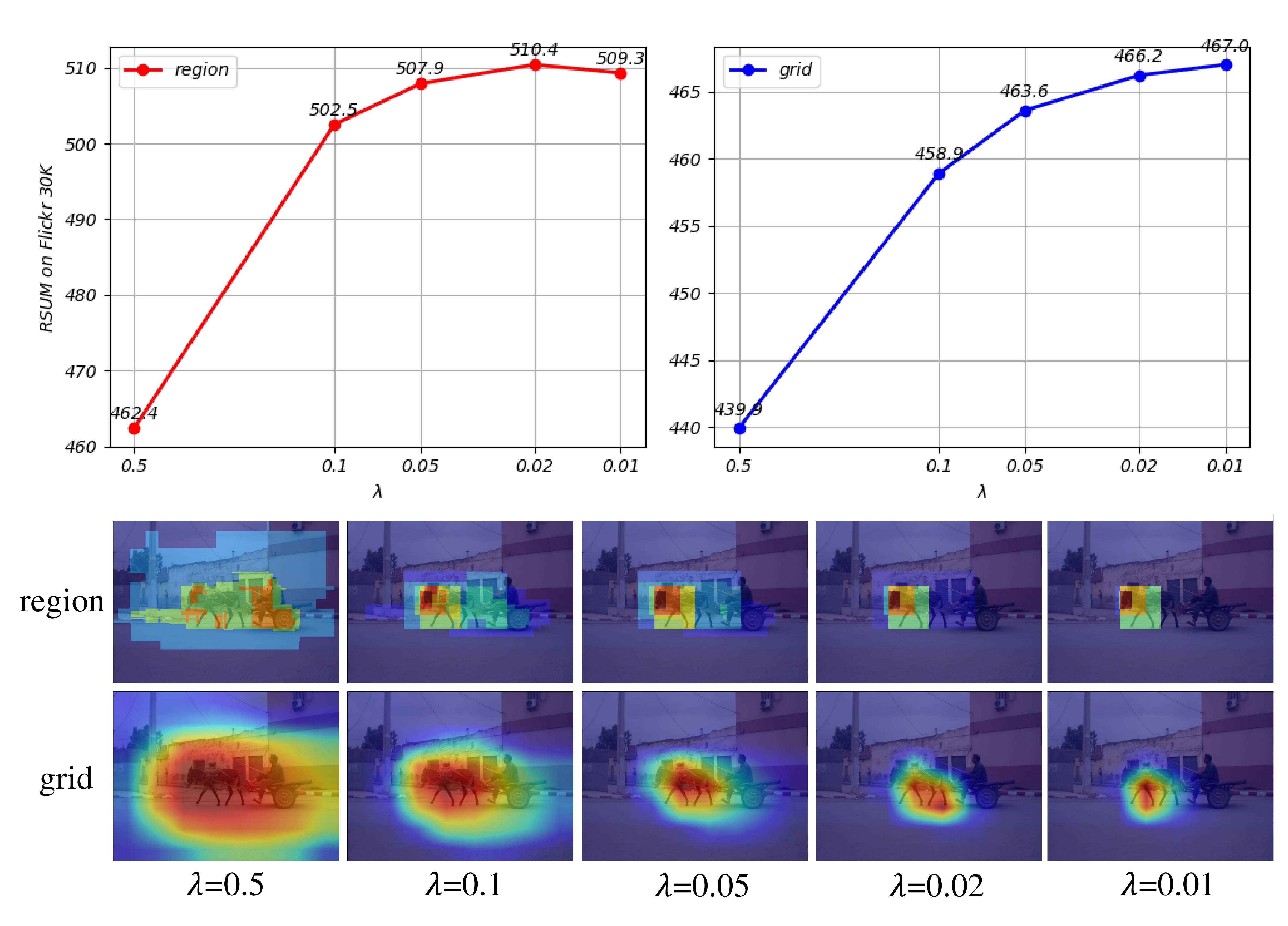}
    \caption{Upper: Performance curve on FLickr30K of hyper-parameters $\lambda$; Lower: Visualized comparison between column-wise transport plans under different $\lambda$.}
    \label{fig:lambda}
\end{figure}
\begin{figure}[tbp]
    \centering
    \includegraphics[width=1\linewidth]{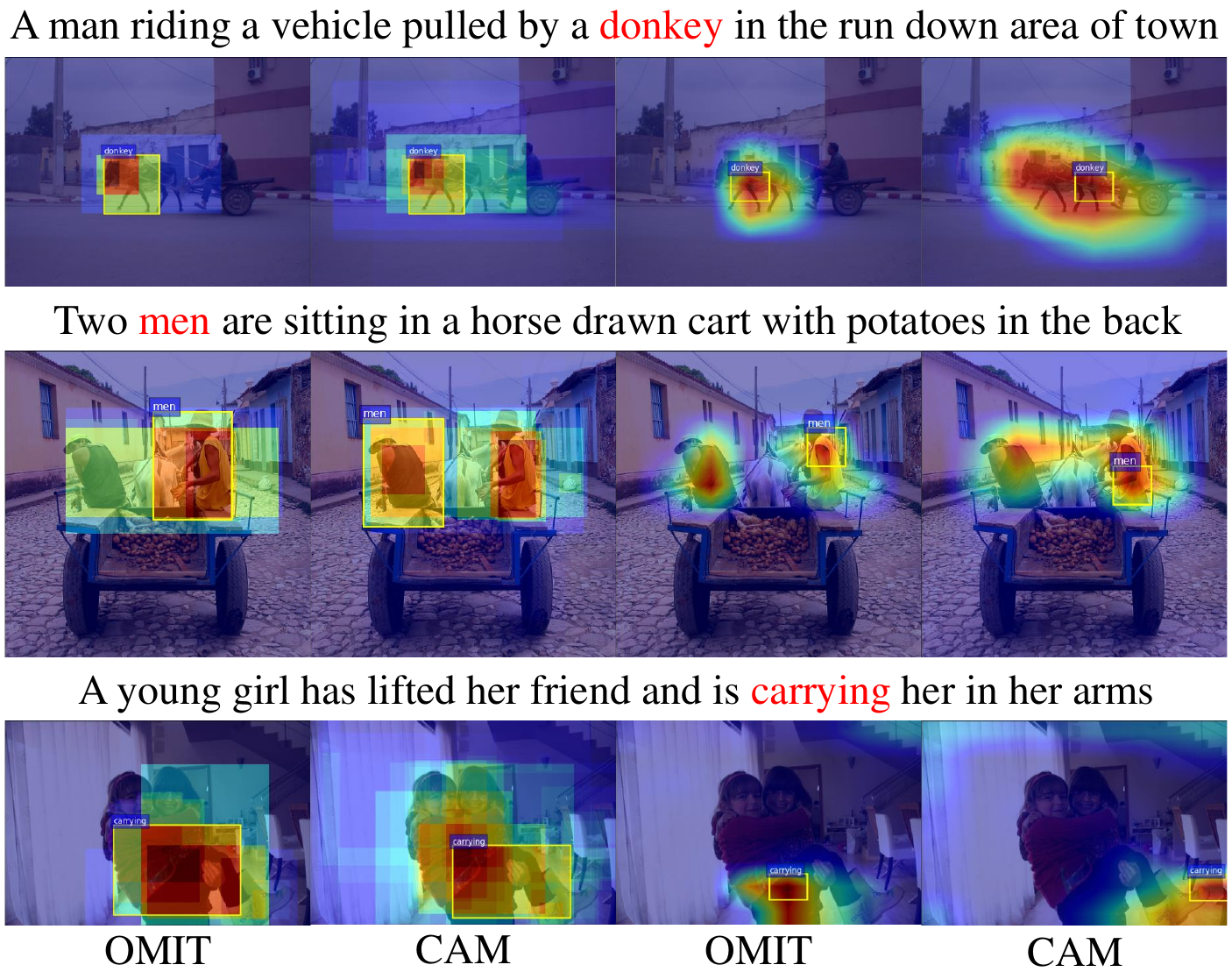}
    \caption{Visual comparisons of alignments between Our OMIT and CAM.}
    \label{fig:omit&cam}
\end{figure}

\mypara{Analysis of Hyper-parameter $\lambda$.} We conduct ablation studies to investigate the impact of the hyper-parameter $\lambda$ defined in Equation~\ref{eq:dsink}. The upper part of Figure~\ref{fig:lambda} illustrates that as $\lambda$ decreases, Sinkhorn distance approaches the CMD, leading to a notable improvement in performance that eventually converges to the optimal result. This finding demonstrates that CMD is an effective metric for capturing semantic similarity. 

Furthermore, we present a visualization of the optimal transport in the lower part of Figure~\ref{fig:lambda}, which offers insights into the contribution of the alignment between the word "donkey" and other visual fragments to the overall similarity. Remarkably, as $\lambda$ decreases, the activated areas become smaller, indicating that "donkey" is correlated with only a minimal amount of granular visual information necessary for establishing its semantic association. This observation supports the inherent sparsity of the optimal plan, as previously highlighted in~\cite{rubner2000earth}. It also aligns with previous findings in~\cite{zhang2022negative}, which suggest that a word is typically associated with a limited number of visual fragments, while the remaining fragments are redundant. 

\mypara{Analysis of the Partial Matching.} We conduct ablation experiments to evaluate the effectiveness of partial matching in OMIT, as summarized in Table~\ref{tab:cmpr_pt}. For clarity of comparison, we present two additional partial matching variants:
\begin{itemize}
    \item "Full": This approach involves directly incorporating global features for full matching.
    \item "Prompt": Motivated by the prompt learning mechanism~\cite{brown2020language, zhou2022learning}, we explore the utilization of two learnable dustbins for partial matching.
\end{itemize}
Surprisingly, partial matching is not only effective for OT but also beneficial for other discrete matching methods like CAM, demonstrating the necessity of reducing redundancy. Compared to the CAM and naive OT baselines, partial matching achieves better performance for both region and grid features, illustrating its robustness across different feature types. Compared to the naive OT approach, the direct addition of global features for matching does not yield noticeable improvements, as it fails to provide additional information. However, the performance of the prompt-based dustbins is not significantly superior, which could be attributed to the absence of clear supervisory signals for effectively learning an optimal threshold. In our experiments, only the partial matching method demonstrates consistently favorable results. Furthermore, improvements are observed in both text retrieval and image retrieval, further demonstrating the effectiveness of partial matching in eliminating redundant alignments.
\begin{table}[tbp]
    \centering
    \small
    \caption{Comparisons between the baselines~(CAM and na\"ive OT) and their corresponding improved versions using partial matching~($+$ Partial) on the Flickr30K test set.}
    \label{tab:cmpr_pt}
    {
        \tabcolsep 2 pt
	    \begin{tabular}{@{\;} c c cccccc@{\;}}
	        \addlinespace
	        \toprule
	        \multicolumn{1}{c}{\multirow{2}{*}{\symtext{TYPE}}} &  \multicolumn{3}{c}{\textsc{img}~$\rightarrow$~\textsc{text}} &      \multicolumn{3}{c}{\textsc{text}~$\rightarrow$~\textsc{img}} & \multicolumn{1}{c}{\multirow{2}{*}{\symtext{rsum}$\uparrow$}} \\ 
	        \cmidrule(lr){2-4} \cmidrule(lr){5-7}
	        {} & \symtext{r}{@1} & \symtext{r}{@5} & \symtext{r}{@10} & \symtext{r}{@1} & \symtext{r}{@5} & \symtext{r}{@10} & \\
                \midrule
                \multicolumn{8}{@{\;}l}{\bf {ResNet-152 \cite{he2016deep} $+$ BiGRU~\cite{chung2014empirical}}} \\[1.5pt]
                CAM & 64.9 & 88.5 & 94.5 & 50.1 & 78.7 & 86.1 & 462.8 \\
                \rowcolor{Gray}Partial CAM & 67.4 & 91.1 & 95.3 & 50.5 & 78.5 & 86.8 & 469.7 \\
			  na\"ive OT & 66.4 & 89.6 & 94.1 & 51.9 & 79.3 & 86.6 & 468.0 \\
                Full & 65.5 & 89.2 & 94.0 & 51.9 & 79.1 & 86.5 & 466.2 \\
                Prompt & 66.7 & 89.4 & 93.6 & 51.0 & 78.7 & 86.2 & 465.5 \\
			  \rowcolor{Gray} Partial OT & 67.6 & 90.6 & 95.2 & 52.5 & 79.3 & 86.6 & \first{471.8} \\
                \midrule
	        \multicolumn{8}{@{\;}l}{\bf {BUTD~\cite{anderson2018bottom} $+$ BiGRU~\cite{chung2014empirical}}} \\[3pt]
                CAM & 79.1 & 95.0 & 97.1 & 58.0 & 84.1 & 90.2 & 503.4 \\
                 \rowcolor{Gray}Partial CAM & 80.5 & 95.3 & 97.7 & 58.7 & 84.2 & 90.3 & 506.8 \\
		      na\"ive OT & 79.4 & 95.6 & 97.8 & 60.9 & 85.5 & 90.7 & 509.9 \\
                Full & 79.0 & 95.1 & 98.1 & 61.1 & 85.5 & 91.0 & 509.9 \\
                Prompt & 79.5 & 95.0 & 98.5 & 61.2 & 85.1 & 90.7 & 510.0 \\
                \rowcolor{Gray}Partial OT & 82.0 & 95.0 & 97.8 & 62.2 & 86.2 & 91.3 & \first{514.5} \\
	       \bottomrule
	    \end{tabular}
	}
\end{table}

\mypara{Compared With Different Margin Initialization Strategies.} In light of the observed disparity in marginal conditions between $\mathcal{S}_{CAM}$ and $\mathcal{S}_{OT}$, we delve into the performance analysis of OMIT under various margin initialization strategies. We consider four different types of margins motivated by current literature: "Uni"~\cite{peyre2019computational}, "Intra"~\cite{liu2019focus}, "Inter"~\cite{zou2022tokenflow} and "Norm"~\cite{yokoi2020word}. In our main text, we have already adopted the uniform margins denoted as "Uni". Here, we provide a brief explanation of each type of margin initialization:
\begin{itemize}
    \item "Uni":~Uniform margins.
    \item "Intra":~Margins initialized based on the proximity of each fragment embedding to their respective intra-modal global embedding.
    \item "Inter":~Margins initialized based on the proximity of each fragment embedding to their inter-modal global embedding.
    \item "Norm":~Margins initialized based on the relative normalization value of the embeddings.
\end{itemize}
 Mathematically, these margins are defined as follows:
\allowdisplaybreaks[4]
\begin{align}
    \text{\textbf{Uni:}}~ \valpha_i=\frac{1}{K_{m}}&,~\vbeta_j=\frac{1}{L_{n}} \notag \\
    \text{\textbf{Intra:}}~ \valpha_i=\frac{\mathop{exp}(\overline{\vct{v}}^{\top} \vct{v}_i/\tau)}{\sum_{i}^{K_{m}}\mathop{exp}(\overline{\vct{v}}^{\top} \vct{v}_i/\tau)}&,~\vbeta_j=\frac{\mathop{exp}(\overline{\vt}^{\top} \vt_j/\tau)}{\sum_{j}^{L_{n}}\mathop{exp}(\overline{\vt}^{\top} \vt_j/\tau)} \notag  \\
    \text{\textbf{Inter:}}~ \valpha_i=\frac{\mathop{exp}(\overline{\vt}^{\top} \vct{v}_i/\tau)}{\sum_{i}^{K_{m}}\mathop{exp}(\overline{\vt}^{\top} \vct{v}_i/\tau)}&,~\vbeta_j=\frac{\mathop{exp}(\overline{\vct{v}}^{\top} \vt_j/\tau)}{\sum_{j}^{L_{n}}\mathop{exp}(\overline{\vct{v}}^{\top} \vt_j/\tau)} \notag  \\
    \text{\textbf{Norm:}}~ \valpha_i = \frac{ \| \widetilde{\vct{v}_i} \|_2}{\sum_{i}^{K_{m}}{\|\widetilde{\vct{v}_i} \|_2}},~ \vbeta_j &= \frac{\|\widetilde{\vt_j}\|_2}{\sum_{i}^{L_{n}}{\|\widetilde{\vt_j}\|_2 }}
    \centering
    \label{eq:margins}
\end{align}

Here, $\vct{v}_i$ and $\vt_j$ denote one of the fragment embeddings. $\widetilde{\vct{v}_i}$ and $\widetilde{\vt_j}$ are fragment embeddings before length normalization, they satisfy $\vct{v}_i = \frac{\widetilde{\vct{v}_i}}{\|\widetilde{\vct{v}_i}\|}$ and $\vt_j = \frac{\widetilde{\vt_j}}{\|\widetilde{\vt_j}\|}$. $\tau$ is the temperature parameter. Based on empirical observations, we set $\tau=1$ in Equation \ref{eq:margins} since we have found that using a smaller value for $\tau$ leads to degraded performance.

The results in Table~\ref{tab:cmpr_mrg} indicate that the choice of different marginal types has limited impact on the performance of the Cross-modal Mover's Distance (CMD), highlighting its robustness within certain limits. Specifically, among the various marginal types examined, the "Intra" variant consistently performs the worst in both benchmarks. This observation suggests that relying solely on the relative proximity within a modality is an unreliable approach for calculating CMD. Conversely, the uniform margins exhibit competitive performance compared to the "Inter" and "Norm" variants. We attribute this success to the assumption of equal contributions from intra-modal fragments, which serves as a significant prior in CMD. 
\begin{table}[tbp]
    \centering
    \small
    {
        \tabcolsep 4 pt
	    \begin{tabular}{@{\;} l c cccccc@{\;}}
	        \addlinespace
	        \toprule
	        \multicolumn{1}{c}{\multirow{2}{*}{\symtext{TYPE}}} &  \multicolumn{3}{c}{\textsc{img}~$\rightarrow$~\textsc{text}} &      \multicolumn{3}{c}{\textsc{text}~$\rightarrow$~\textsc{img}} & \multicolumn{1}{c}{\multirow{2}{*}{\symtext{rsum}$\uparrow$}} \\ 
	        \cmidrule(lr){2-4} \cmidrule(lr){5-7}
	        {} & \symtext{r}{@1} & \symtext{r}{@5} & \symtext{r}{@10} & \symtext{r}{@1} & \symtext{r}{@5} & \symtext{r}{@10} & \\
                \midrule
                \multicolumn{8}{@{\;}l}{\bf {ResNet-152 \cite{he2016deep} $+$ BiGRU~\cite{chung2014empirical}}} \\[1.5pt]
                \rowcolor{Gray} Uni & 67.0 & 90.3 & 95.6 & 51.8 & 79.4 & 87.1 & \first{471.2} \\
                Intra & 65.4 & 88.6 & 93.9 & 51.5 & 78.6 & 86.2 & 464.3 \\
                Inter & 67.7 & 89.3 & 94.8 & 51.9 & 79.5 & 86.9 & 470.1 \\
                Norm & 68.2	& 88.8 & 94.2 & 52.2 & 79.8	& 87.4 & 470.6 \\
                \midrule
	        \multicolumn{8}{@{\;}l}{\bf {BUTD\cite{anderson2018bottom} $+$ BiGRU~\cite{chung2014empirical}}} \\[3pt]
		      \rowcolor{Gray} Uni & 78.8 & 95.4 & 98.0 & 60.6 & 85.1 & 91.1 & 509.0 \\
                Intra & 78.3 & 95.4 & 97.8 & 59.7 & 85.1 & 90.8 & 507.1 \\
                Inter & 78.9 & 95.4 & 98.3 & 59.9 & 84.8 & 91.0 & 508.3 \\
                Norm & 80.0	& 95.9 & 97.4 & 59.6 & 85.2 & 91.0 & \first{509.1} \\
	       \bottomrule
	    \end{tabular}
	}
    \caption{Comparisons of OMIT variants with different marginal conditions on the Flickr30K test set.}
    \label{tab:cmpr_mrg}
\end{table}
\begin{figure*}[htbp] 
    \centering
    \captionsetup{aboveskip=2pt, belowskip=0pt} 
    {\centering
    \Large\textit{\textbf{Query}: A man riding a vehicle pulled by a donkey in the run down area of town.}\par
    }
    \begin{subfigure}[b]{0.4995\linewidth}
        \centering
        \includegraphics[width=\linewidth]{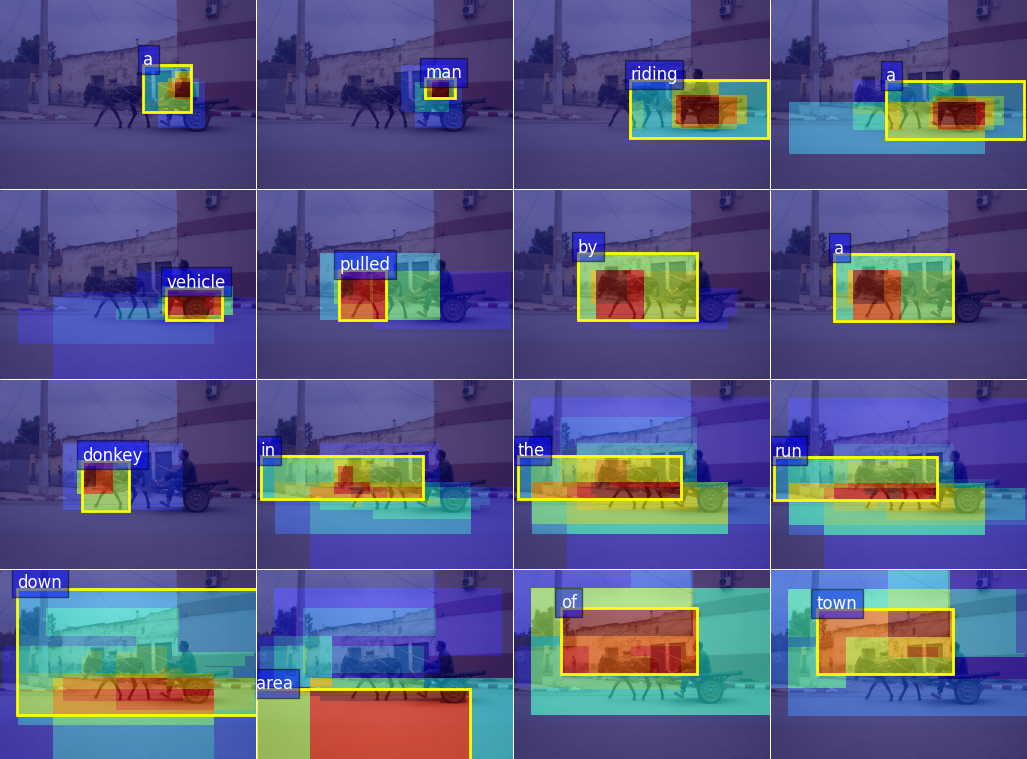}
        \label{fig:supp_figure4_donkey_region}
    \end{subfigure}
    \begin{subfigure}[b]{0.495\linewidth}
        \centering
        \includegraphics[width=\linewidth]{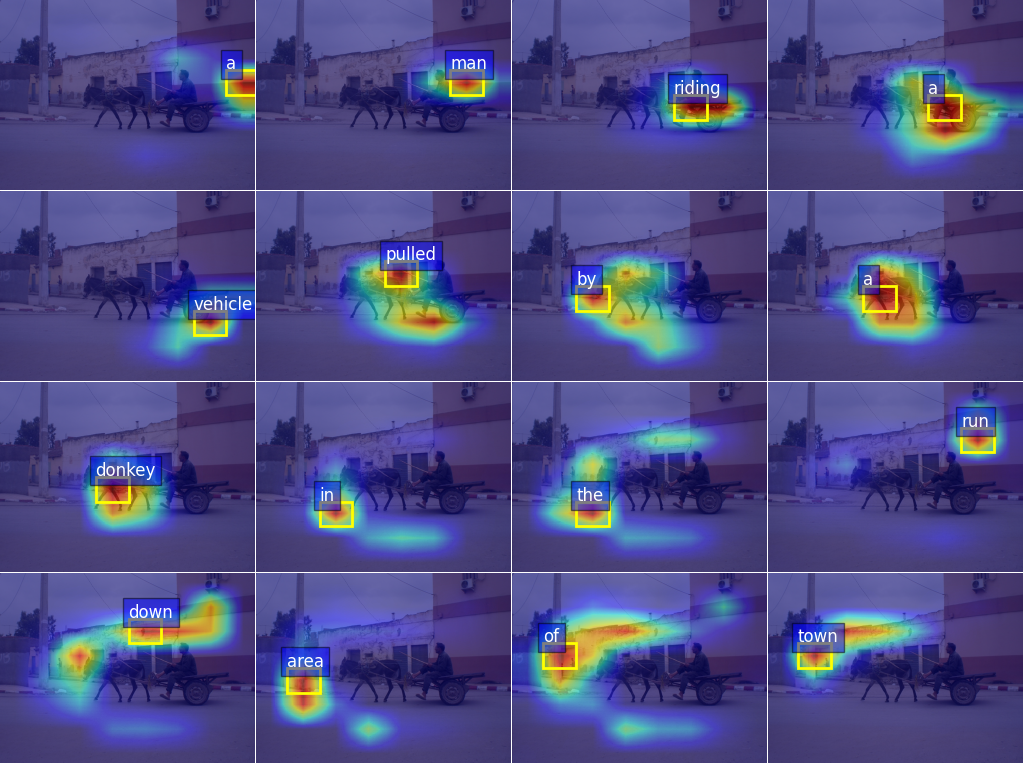}
        \label{fig:supp_figure4_donkey_grid}
    \end{subfigure}
    \caption{An illustrative example of image-text matching showcasing the column-wise transport plan generated by our OMIT variants. Left: region-based, Right: grid-based.}
    \label{fig:supp_figure4_donkey_combined}
\end{figure*}

\subsection{Visualization}
\begin{figure}[tbp]
    \centering
    \includegraphics[width=1\linewidth]{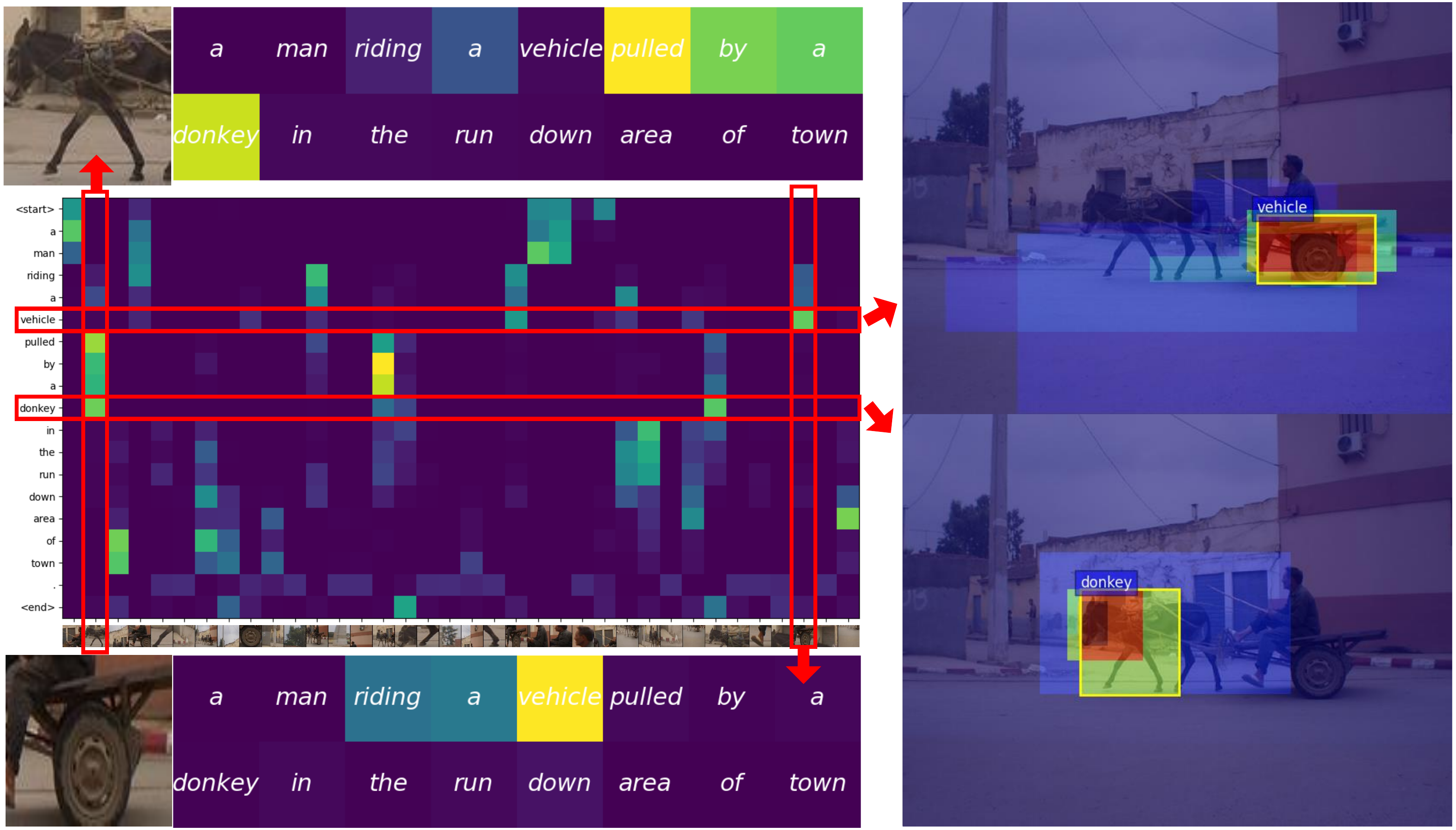}
    \caption{The visualized optimal transport plan with its row-wise and column-wise visualizations. }
    \label{fig:visualization}
\end{figure}

Figure~\ref{fig:supp_figure4_donkey_combined}-\ref{fig:visualization} provides some visualizations to help understand the optimal transport plan. We visualize two row-wise alignments, which depict the relationships between a word and all regions, as well as two column-wise alignments, illustrating the connections between a region and all words. The visualizations clearly demonstrate that each word or region selectively corresponds to a sparse set of meaningful regions or words, while disregarding irrelevant candidates. Notably, the words "donkey" and "vehicle" exhibit strong associations with consistent regions, and the regions representing donkey and vehicle also align well with the words "donkey," "vehicle," and their associated prefix words. These visualizations provide valuable insights into the alignment patterns of optimal transport plan that capture meaningful associations while disregarding irrelevant information.



\section{Conclusion}
\label{sec:5-conclu}
In this work, we present crOss-Modal sInkhorn maTching~(OMIT), which aligns fine-grained visual and semantic fragments by leveraging Optimal Transport. OMIT offers several advantages over existing matching methods. Firstly, OMIT capitalizes on granular information and explicitly establishes associations between cross-modal fragments, leading to superior performance compared to VSE and PEM. Secondly, by utilizing the Sinkhorn distance as the similarity metric, OMIT surpasses the performance of current SOTAs due to its inherent focal matching. Thirdly, OMIT bridges the efficiency gap between global embedding methods and current local matching approaches thanks to efficient Bregman iterations. By incorporating the technique of partial matching, OMIT , achieving the best results on benchmarks, demonstrating its potential. We foresee that future research in fine-grained cross-modal pre-training can be built upon the foundations established by OMIT.

\clearpage
\begin{appendix}
\newcommand{\beginsupplementA}{%
    \setcounter{table}{0}
    \renewcommand{\thetable}{A\arabic{table}}%
    \setcounter{figure}{0}
    \renewcommand{\thefigure}{A\arabic{figure}}%
    \setcounter{algorithm}{0}
    \renewcommand{\thealgorithm}{A\arabic{algorithm}}%
    \setcounter{section}{0}
    \renewcommand{\thesection}{A\arabic{section}}%
}
\beginsupplementA
\section{Supplements to the Method}
\subsection{Revisiting Current Methods}
The underlying principle of cross-modal matching revolves around associating each individual data sample, such as an image or a sentence, with a distribution in a shared space. Within this shared space, diverse semantics are demarcated by distinct regions of the distribution, effectively capturing semantic variations. One approach to construct such a distribution involves considering partial semantics as sampling points. This results in a discrete distribution that encapsulates the essence of data sample. To illustrate, consider an image denoted as $\mat{\calV}_m = \{\vv_i|_{i=1}^{K_m}\}$ and a caption as $\mat{\calT}_n = \{\vt_j|_{j=1}^{L_n}\}$. Here, $m$ and $n$ index the dataset, $\vv_i \in \R^d$ and $\vt_j \in \R^d$ denote the image sub-regions/patches or text tokens, and $K_m$ and $L_n$ signify the number of fragments. The formulation of the discrete distribution $\mat{\calV}_m$ and $\mat{\calT}_n$ is as follows:
\begin{equation}
    \centering
    p(\mat{\calV}_m) \coloneqq \sum_{i=1}^{K_m}\valpha_i\delta_{\vv_i},~ p(\mat{\calT}_n) \coloneqq \sum_{j=1}^{L_n}\vbeta_j\delta_{\vt_j}
    \label{eq:scan}
\end{equation}
Here, $\delta_x$ represents the Dirac function centered at location $x$, symbolizing a unit of mass. $\valpha\in\R^{K_m}$ and $\vbeta\in\R^{L_n}$ are discrete probability vectors that sum to 1. Typically, we set $\valpha_i=\frac{1}{K_{m}}$ and $\vbeta_j=\frac{1}{L_{n}}$. The primary challenge in cross-modal matching lies in quantifying the similarity and effectively capturing the semantic relationship between $p(\mat{\calV}_m)$ and $p(\mat{\calT}_n)$.

In the subsequent section, we present a succinct overview of the prevailing methods, specifically Visual-Semantic Embedding (VSE), Cross-Attention-based Method (CAM), and Probabilistic Embedding Method (PEM). This comprehensive analysis endeavors to provide a panoramic understanding and further insights.

\mypara{VSE.}~The VSE methodology, while straightforward, inadvertently overlooks the high-order statistics of $p(\mat{\calV}_m)$ and $p(\mat{\calT}_n)$. Rather than considering these statistics, it relies solely on the cosine similarity between the average vectors of discrete distributions to estimate their similarity. However, this approach's limitation becomes apparent in its failure to account for high-order statistics, resulting in a lack of semantic diversity and limited effectiveness in handling ambiguity. Given that fragment embeddings are typically normalized to the unit sphere, VSE essentially computes the average cosine similarity across all pairs of cross-modal fragments:
\begin{align}
    \centering
    \begin{split}
        \mathcal{S}_{VSE} &\coloneqq \frac{(\frac{1}{K_m}\sum_{i=1}^{K_m}\vv_i)^\top (\frac{1}{L_n}\sum_{j=1}^{L_n}\vt_j)}{\|\frac{1}{K_m}\sum_{i=1}^{K_m}\vv_i\|_2 \cdot \|\frac{1}{L_n}\sum_{j=1}^{L_n}\vt_j\|_2}  \\
        &= \frac{\frac{1}{K_mL_n}\sum_{i=1}^{K_m}\sum_{j=1}^{L_n} \vv_i^\top \vt_j}{\| \overline{\vv} \|_2 \cdot \| \overline{\vt} \|_2} \\
        &\stackrel{\textcircled{1}}{\geq} \sum_{i=1}^{K_m}\sum_{j=1}^{L_n}\frac{1}{K_m L_n}\vv_i^\top \vt_j
    \end{split}
    \label{eq:dvse}
\end{align}

In the above equation, the terms $\overline{\vv}=\frac{1}{K_{m}}\sum_{i=1}^{K_{m}}{\vv_i}$ and $\overline{\vt}=\frac{1}{L_{n}}\sum_{j=1}^{L_{n}}{\vt_j}$ represent the average (or global) embedding of $\mat{\calV}_m$ and $\mat{\calT}_n$, respectively. The validity of $\textcircled{1}$ can be attributed to the triangle inequality, which states that $\| \frac{1}{K_{m}}\sum_{i=1}^{K_{m}} \vv_i \|_2 \leq \frac{1}{K_{m}}\sum_{i=1}^{K_{m}}{\| \vv_i \|_2}=1 $, similar to $\|\overline{\vt}\|_2\leq1$. However, it should be noted that the uniform treatment of all pairwise alignments in VSE may pose challenges, as certain alignments might contribute minimally to the overall semantic coherence.

\mypara{CAM.} CAM represents an endeavor to achieve more interpretable matching results in contrast to VSE, by leveraging an attention mechanism to focus on well-aligned correspondences. The modification introduced in CAM can be observed in Equation~\ref{eq:dcam}, where the average operation is replaced with a weighted summation, as illustrated below:
\begin{align}
    \centering
        \mathcal{S}_{CAM} \coloneqq \sum_{i=1}^{K_m}\sum_{j=1}^{L_n}\omega_{ij}\cdot\vv_i^\top \vt_j
    \label{eq:dcam}
\end{align}
where $\omega_{ij}$ signifies the significance of the alignment $\vv_i^\top \vt_j$, which usually obeys:
\begin{equation}
    \begin{split}
        \omega_{ij} &\coloneqq \frac{\eta_{ij}}{\|\sum_{i=1}^{K_m}{\eta_{ij}\vv_i} \|_2} \stackrel{\textcircled{2}}{\geq} \eta_{ij} \\
        \text{where}~~\eta_{ij} &= \frac{1}{L_n}\frac{\mathop{exp}(\vv_i^\top \vt_j/\tau)}{\sum_{i=1}^{K_m}\mathop{exp}(\vv_i^\top \vt_j/\tau)}
    \end{split}
    \centering
    \label{eq:omega}
\end{equation}

Here, the temperature parameter $\tau$ is incorporated into the softmax function, as outlined by~\cite{hinton2015distilling}. $\textcircled{2}$ holds true also due to the triangle inequality. Notably, several works~\cite{lee2018stacked,zhang2022negative,pan2023fine} have empirically discovered that selectively disregarding certain alignments can enhance the discriminative power of $\mathcal{S}_{CAM}$. This observation is realized through the following equation:
\begin{equation}
    \begin{split}
        \eta_{ij} &= \frac{1}{L_n}\frac{\mathop{exp}(\vv_i^\top \vt_j/\tau)}{\sum_{\vv_i\in\mat{\Delta}}\mathop{exp}(\vv_i^\top \vt_j/\tau)} \\
        \text{where}~~\mat{\Delta} &= \{ \vv_k | \vv_k^\top \vt_j \geq\epsilon,~\forall k\in [1,K_m] \}
    \end{split}
    \centering
    \label{eq:sparsity}
\end{equation}

In the above equation, the sparse threshold $\epsilon$ varies depending on the specific methods employed. 

The CAM methodology bestows greater gradient emphasis upon visual fragments that are distanced from the corresponding text tokens. This results in a more discerning representation when juxtaposed with VSE. However, CAM faces a distinct challenge of redundant alignments, which diminishes its overall effectiveness. Another noteworthy aspect is the asymmetry of $\omega_{ij}$ in t2i~(text-to-image) attention and i2t~(image-to-text) attention, a characteristic present in the original CAM formulation~\cite{lee2018stacked}. This asymmetry introduces complexities when attempting to transfer the CAM framework to other tasks.

\mypara{PEM.} PEM considers both first-order and second-order statistics. It model the covariance of fragment embeddings and describe the data holistically using parametric Gaussian distributions. These methods utilize commonly used metrics such as $\mathop{KL}$-divergence, $\mathop{JS}$-divergence, and Wasserstein distance quantify the dissimilarity between continuous distributions. Assuming $p(\mat{\calV}_m) \sim \mathcal{N}(\vmu_{\mat{\calV}}, \mSigma_{\mat{\calV}})$ and $p(\mat{\calT}_n) \sim \mathcal{N}(\vmu_{\mat{\calT}}, \mSigma_{\mat{\calT}})$ where $\mathcal{N}(\vmu, \mSigma)$ denotes an isotropic Gaussian distribution with the mean vector $\vmu$ and covariance matrix $\mSigma$, the discrepancy measured by the Wasserstein distance is shown in Equation \ref{eq:dpem}:
\begin{align}
    \mathcal{D}_{PEM} &\coloneqq \mathop{min}\limits_{\vgamma \in \mGamma(p(\mat{\calV}_m),p(\mat{\calT}_n))} \sqrt{\int_{(\vv,\vt)\sim\vgamma} \|\vv-\vt\|^2_2 d\vgamma(\vv,\vt)} \notag \\
    &= \sqrt{\|\vmu_{\mat{\calV}}-\vmu_{\mat{\calT}}\|^2_2+\|\mSigma_{\mat{\calV}}-\mSigma_{\mat{\calT}}\|^2_2})
    \centering  
    \label{eq:dpem}
\end{align}
where $\mGamma(p(\mat{\calV}_m),p(\mat{\calT}_n))$ denotes the sets of all joint distributions $\vgamma(\vv,\vt)$ whose marginals are $p(\mat{\calV}_m)$ and $p(\mat{\calT}_n)$, respectively. Intuitively, $\vgamma(\vv,\vt)$ indicates how much “mass” must be transported from $\vt$ to $\vv$ in order to transform the distributions $p(\mat{\calT}_n)$ into the distribution $p(\mat{\calV}_m)$. We argue that the improvement of PEM comes from the robust training by introducing uncertainty. However, the assumption of a parametric distribution may not be essential.

\subsection{Methodology Details about Optimal Transport}
In this subsection, we provide further details on the optimization of entropic regularized Optimal Transport. Recall the source problem
\begin{align}
    \begin{split}
        \mathcal{D}_{OT} &= \mathop{min}\limits_{\mOmega\in\mU(\valpha,\vbeta)} \left\langle \mOmega,\mC \right\rangle\coloneqq\sum_{i=1}^{K_m}\sum_{j=1}^{L_n}\omega_{ij}c_{ij} \\
        \text{subject to}~~&\mOmega\ones_{L_n} = \valpha,~\mOmega^\top\ones_{K_m} = \vbeta,~\mOmega\in \R_+^{K_m\times L_n} \\
    \end{split}
    \label{eq:dot_ap}
    \centering
\end{align}
and its entropic regularized version:
\begin{equation}
    \centering
    \begin{split}
            \mathcal{D}_{S} &= \mathop{min}\limits_{\mOmega\in\mU(\valpha,\vbeta)} \left\langle \mOmega,\mC \right\rangle -\lambda h(\mOmega) \\
        &\coloneqq \sum_{i=1}^{K_m}\sum_{j=1}^{L_n}\omega_{ij}c_{ij}+\lambda\omega_{ij}(\log\omega_{ij}-1)
    \end{split}
    \label{eq:dsink_ap}
\end{equation}
where $h(\mOmega)=-\sum_{i=1}^{K_m}\sum_{j=1}^{L_n}\omega_{ij}(\log \omega_{ij}-1)$ is the entropic term; $\omega_{ij}$ and $c_{ij}$ refer to the element in the $i$-th row and $j$-th column of the plan matrix $\mOmega$ and cost matrix $\mC$, respectively. By introducing entropic regularization, the source problem is transformed into a p-strongly convex problem, which can be efficiently solved using Lagrange multipliers:
\begin{equation}
    \centering
    \begin{split}
        \mathcal{J}(\mOmega, \vmu, \vtheta) =& \left\langle \mOmega,\mC \right\rangle -\lambda h(\mOmega) \\
        -\vmu^{\top}(\mOmega\ones_{L_n}-\valpha)&-\vtheta^{\top}(\mOmega^\top\ones_{K_m}-\vbeta)
    \end{split}
    \label{eq:E}
\end{equation}
where $\vmu\in\R^{K_m}, \vtheta\in\R^{L_m}$ are the dual variables for the two equality constraints in $\mU(\valpha,\vbeta)$. For any couple $(i, j)$, When $\frac{\partial\mathcal{J}}{\partial\omega_{ij}}=c_{ij}+\lambda\log(\omega_{ij})-\mu_{i}-\theta_{j}=0$, Equation~\ref{eq:dsink_ap} obtains the unique solution $\mOmega^*$ whose every element obeys the form $\omega_{ij}=e^{\frac{\mu_i}{\lambda}}e^{-\frac{c_{ij}}{\lambda}}e^{\frac{\theta_j}{\lambda}}$. To this end, $\mOmega^*$ has the form:
\begin{equation}
    \centering
    \mOmega^* = \mathop{diag}(\vmu^{(t)})\mxi\mathop{diag}(\vtheta^{(t)})
    \label{eq:lemma}
\end{equation}
where $\mxi=e^{-\frac{\mC}{\lambda}}$ is the Gibbs kernel as previously stated in the main text, $\vmu^{(t)}$ and $\vtheta^{(t)}$ are the projected vector to be calculated, which can be obtained by transforming $\vmu$ and $\vtheta$. Directly seeking the analytical solutions of $\vmu$ and $\vtheta$ can be complicated; hence, iterative optimization is a better approach. According to~\cite{peyre2019computational}, current optimization solutions of Equation~\ref{eq:dsink_ap} are mainly from two perspectives, namely matrix scaling~\cite{cuturi2013sinkhorn} and alternating projection~\cite{benamou2015iterative}. Both are essentially equivalent, and we adopt the latter method in our main text. 

\begin{figure*}[htbp]
    \centering
    \includegraphics[width=1\linewidth]{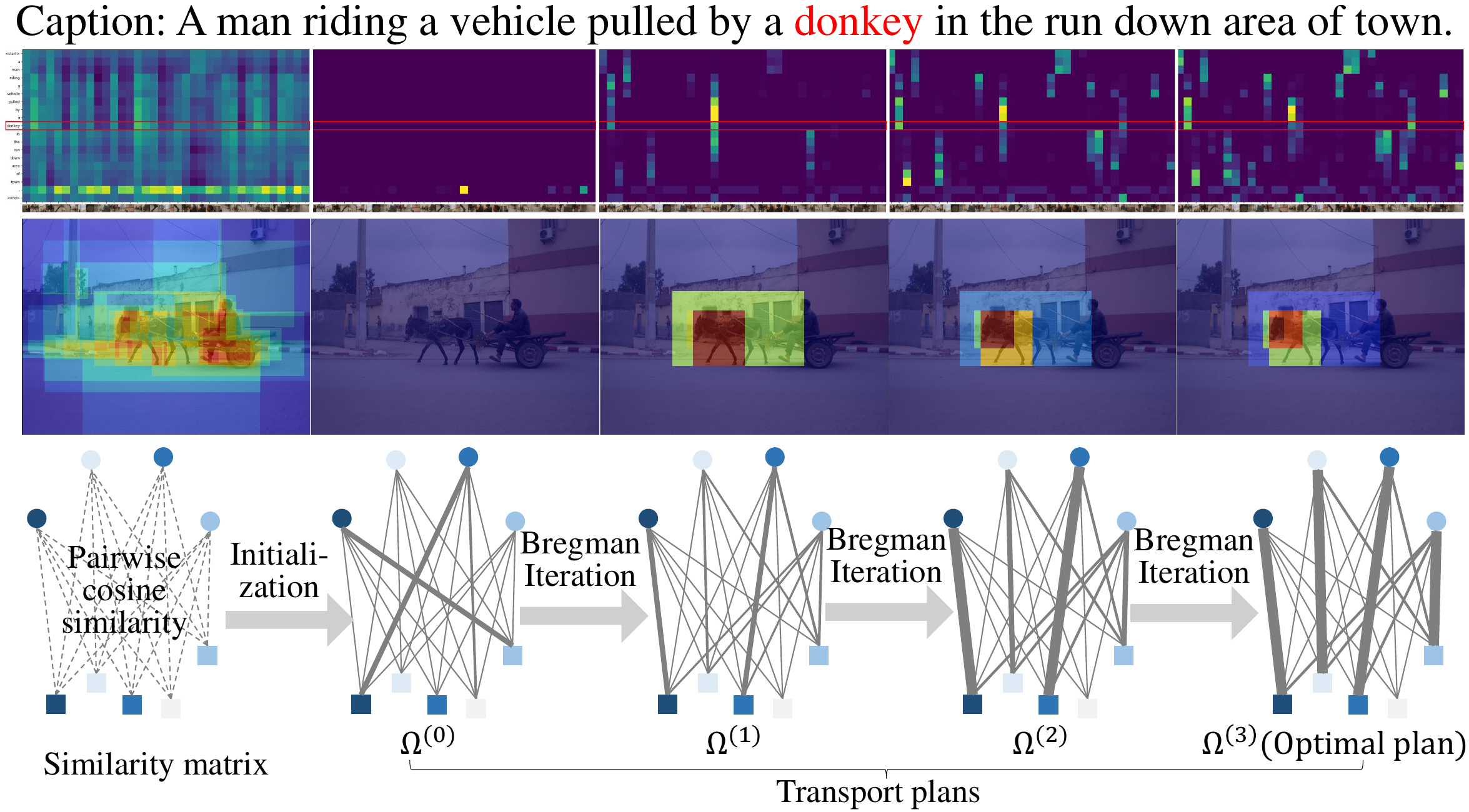}
    \caption{An illustration of the Sinkhorn matching module which is elaborated in the main text. Each column corresponds to a time step, and the three rows from top to bottom respectively depict the changes over iterations in the overall transport plan, the row-wise transport plan and the demo sketch. It's evident that the transport plan gradually becomes sparser as the iterations process.}
    \label{fig:sinkhorn}
\end{figure*}

\renewcommand{\algorithmicrequire}{\textbf{Input:}}
\renewcommand{\algorithmicensure}{\textbf{Output:}}
\begin{algorithm}[htb]
	\caption{Leveraging Bregman iterations to optimize the Sinkhorn distance between an image and a caption.}
	\label{alg:a1}
	\begin{algorithmic}[1]
            \REQUIRE Fragment embeddings of an image $\mat{\mathcal{V}}_m\in\R^{K_m\times d}$\cmmnt{$\{\vv_i|_{i=1}^{K_m}\}$}, fragment embeddings of a caption $\mat{\mathcal{T}}_n\in\R^{L_n\times d}$, hyper-parameter~$\lambda$.
            \ENSURE Sinkhorn similarity $\mathcal{S}_{OT}$.
		\STATE Calculate the cost matrix:~$\mC=1-\mat{\mathcal{V}}_m\mat{\mathcal{T}}_n^{\top}\in\R^{K_m\times L_n}$.
  		\STATE Calculate the initial transport plan:~$\mOmega^{(0)}=e^{-\frac{\mC}{\lambda}}$.
            \STATE Initialize $\valpha = \frac{1}{K_{m}}\ones_{K_m}$,~$\vbeta = \frac{1}{L_{n}}\ones_{L_n}$, \\
            $\varepsilon=1e-6$, maximum iterations $T=3$.
            \FOR{$t=1,\cdots, T$}
		\STATE Perform row-wise normalization:~$\mOmega^{(t)} \leftarrow \mathop{diag}(\frac{\valpha}{\mOmega^{(t)}\ones_{L_n}})\mOmega^{(t-1)}$
            \STATE Perform column-wise normalization:\\
            $\mOmega^{(t)} \leftarrow \mOmega^{(t)}\mathop{diag}(\frac{\vbeta}{\mOmega^{(t)\top}\ones_{K_m}})$
            \IF{${{{\left\| \mOmega^{(t)} - \mOmega^{(t-1)} \right\|_2} \mathord{\left/{\vphantom {{\left\| \mOmega^{(t)} - \mOmega^{(t-1)}\right\|_2} {\left\| \mOmega^{(t-1)} \right\|_2}}} \right. \kern-\nulldelimiterspace} {\left\| \mOmega^{(t-1)} \right\|_2}}}  < \varepsilon $}
            \STATE{break}
            \ENDIF
            \ENDFOR
            \STATE Calulate the Sinkhorn similarity:~$\mathcal{S}_{OT} = 1-\left\langle \mOmega^{(t)},\mat{\calV}_m\mat{\calT}_n^{\top} \right\rangle$
            \RETURN $\mathcal{S}_{OT}$
	\end{algorithmic}  
\end{algorithm}

\mypara{Matrix scaling.} Specifically, to satisfy the marginal condition of $\mOmega^*$, we combine it with Equation~\ref{eq:lemma}, leading to the following expression:
\begin{equation}
    \centering
    \begin{split}
        \vmu^{(t)}\odot(\mxi\vtheta^{(t)}) = \valpha,~~ \vtheta^{(t)}\odot(\mxi^{\top}\vmu^{(t)}) = \vbeta
    \end{split}
    \label{eq:margin}
\end{equation}
where $\odot$ refers to the Hadamard product. The problem of solving Equation~\ref{eq:margin}, which involves scaling the rows and columns of the matrix $\mK$ to satisfy the given marginal constraints, is widely known in the numerical analysis community as the matrix scaling problem~\cite{nemirovski1999complexity}. This problem can be efficiently solved using the well-known Sinkhorn-Knopp algorithm~\cite{knight2008sinkhorn}, which can be expressed as:
\begin{equation}
    \centering
    \begin{split}
        \vmu^{(t+1)} = \frac{\valpha}{\mxi\vtheta^{(t)}},~~\vtheta^{(t+1)} = \frac{\vbeta}{\mxi^{\top}\vmu^{(t+1)}}
    \end{split}
    \label{eq:sinkhorn}
\end{equation}
The division operator here denotes the Hadamard division. The convergence proof of Equation~\ref{eq:sinkhorn} is attributed to Sinkhorn and Knopp~\cite{sinkhorn1964relationship}, hence the name of the algorithm. The initiation of $\vtheta^{(0)}$ can be arbitrary, such as $\vtheta^{(0)}=\ones_{L_n}$. 

\mypara{Alternating projection.} Equation~\ref{eq:dsink_ap} can be re-cast as follows:
\begin{equation}
    \centering
    \begin{split}
        \mathcal{D}_{S} &= \mathop{min}\limits_{\mOmega\in\mU(\valpha,\vbeta)} \lambda\sum_{i=1}^{K_m}\sum_{j=1}^{L_n}\omega_{ij}(\log\frac{\omega_{ij}}{e^{-\frac{c_{ij}}{\lambda}}}-1) \\
        &= \mathop{min}\limits_{\mOmega\in\mU(\valpha,\vbeta)} \lambda\mathop{KL}(\mOmega|\mxi) \\
        &\coloneqq \lambda\sum_{i=1}^{K_m}\sum_{j=1}^{L_n}(\omega_{ij}\log\frac{\omega_{ij}}{\xi_{ij}}-\omega_{ij}+\xi_{ij})
    \end{split}
    \label{eq:kl_ap}
\end{equation}

If the convex sets $\mU(\valpha,\vbeta)$ are indeed affine subspaces, then Equation~\ref{eq:kl_ap} can be solved simply using na\"ive Bregman projections:
\begin{equation}
    \centering
    \begin{split}
         \mOmega^{*} = {\mathop{Proj}}_{\mU(\valpha,\vbeta)}^{\mathop{KL}}(\mOmega) \coloneqq \mathop{argmin}\limits_{\mOmega\in\mU(\valpha,\vbeta)}\mathop{KL}(\mOmega|\mxi)
    \end{split}
    \label{eq:proj}
\end{equation}
However, denoting $\mathcal{C}_{\valpha}^{1} \coloneqq\{\mOmega\in \R_+^{K_m\times L_n}|\mOmega\ones_{L_n}=\valpha\}$ and $\mathcal{C}_{\vbeta}^{2}\coloneqq\{\mOmega\in \R_+^{K_m\times L_n}|\mOmega^{\top}\ones_{K_m}=\vbeta\}$ the rows and columns constraints, we have $\mU(\valpha,\vbeta) = \mathcal{C}_{\valpha}^{1} \bigcap\mathcal{C}_{\vbeta}^{2}$. It can be shown that the convex set $\mU(\valpha,\vbeta)$ is the intersection of these two affine subspaces, i.e., $\mathcal{C}_{\valpha}^{1}$ and $\mathcal{C}_{\vbeta}^{2}$. Due to this non-trivial intersection, the na\"ive Bregman projections in Equation~\ref{eq:proj} do not converge to the optimal solution. Fortunately, as discussed in~\cite{benamou2015iterative}, this issue can be addressed by extending the Dykstra algorithm~\cite{boyle1986method} to the KL setting, that is:
\begin{equation}
    \centering
    \begin{split}
         \mOmega^{*} &= \mathop{argmin}\limits_{\mOmega\in{\mathcal{C}_{\valpha}^{1} \bigcap\mathcal{C}_{\vbeta}^{2}}}\mathop{KL}(\mOmega|\mxi) \\
         &\coloneqq {\mathop{Proj}}_{\mathcal{C}_{\vbeta}^{2}}^{\mathop{KL}}\left({\mathop{Proj}}_{\mathcal{C}_{\valpha}^{1}}^{\mathop{KL}}(\mOmega)\right)
    \end{split}
    \label{eq:iterproj}
\end{equation}

According to~\cite{benamou2015iterative}, we have:
\begin{equation}
    \begin{split}
        \mOmega^{(2t+1)} & = {\mathop{Proj}}_{\mathcal{C}_{\valpha}^{1}}^{\mathop{KL}}(\mOmega^{(2t)}) \\
        &\coloneqq \mathop{diag}(\frac{\valpha}{\mOmega^{(2t)}\ones_{L_n}})\mOmega^{(2t)} \\
        \mOmega^{(2t+2)} & = {\mathop{Proj}}_{\mathcal{C}_{\vbeta}^{2}}^{\mathop{KL}} (\mOmega^{(2t+1)}) \\
         &\coloneqq \mOmega^{(2t+1)}\mathop{diag}(\frac{\vbeta}{\mOmega^{(2t+1)\top}\ones_{K_m}}) \\
        &\text{Initialize}~~\mOmega^{(0)}=\mxi
    \end{split}
    \centering
    \label{eq:projection_ap}
\end{equation}

To this end, We provide the derivation of solving the entropic regularized Optimal Transport by alternating Bregman projections. Algorithm~\ref{alg:a1} provides an overview of the Bregman iterations for image-text matching. Notably, iterations behind Equation~\ref{eq:sinkhorn} can be viewed as a special case of iterations behind Equation~\ref{eq:projection_ap}, as demonstrated in~\cite{peyre2019computational}. In our main text, we adopt Equation~\ref{eq:projection_ap} to optimize the sinkhorn distance for which we can directly visualize the intermediate transport plan $\mOmega^{(2t)}$ of each $t$.

Furthermore, the similarity between an image $\mat{\calV}_m$ and a caption $\mat{\calT}_n$ using Optimal Transport, as defined in Equation~6 of our main text, can be expressed as:
\begin{equation}
    \centering
    \begin{split}
        \mathcal{S}_{OT} = \left\langle \mOmega^*,\mat{\calV}_m\mat{\calT}_n^{\top} \right\rangle &\coloneqq \sum_{i=1}^{K_m}\sum_{j=1}^{L_n}\omega_{ij}\cdot\vv_i^\top \vt_j \\
        \text{where}~\sum_{j=1}^{L_n}\omega_{ij}=\valpha_i,\sum_{i=1}^{K_m}\omega_{ij}&=\vbeta_j,\sum_{i=1}^{K_m}\sum_{j=1}^{L_n}\omega_{ij}=1
    \end{split}
    \label{eq:sot}
\end{equation}

Combining Equation~\ref{eq:dvse}, \ref{eq:dcam} and \ref{eq:sot} we can find that $\mathcal{S}_{OT}$ serves as an upper bound of $\mathcal{S}_{VSE}$ and $\mathcal{S}_{CAM}$. However, the disparity arises from the fact that the marginal conditions of $\omega$ in $\mathcal{S}_{VSE}$ align with those of $\mathcal{S}_{OT}$, whereas the marginal conditions of $\omega$ in $\mathcal{S}_{CAM}$ do not align. In light of these findings, it is more reasonable to directly compare the bound similarities rather than relying solely on empirical similarities. 

We present a visualization of the Bregman iterations used to solve Equation~\ref{eq:projection_ap} in Figure~\ref{fig:sinkhorn}, which specifies the Sinkhorn Matching module of Figure 2 in the main text. The upper row of Figure~\ref{fig:sinkhorn} depicts the similarity matrix (inverse cost matrix) and transport plans $\mOmega^{(t)}$ at each step $t$ (here $t=0,\cdots,3$). The middle row shows the alignments between the word "donkey" in the caption and all other salient regions in the image, which correspond to the row in $\mOmega^{(t)}$ marked with a red box. The lower row is a sketch illustrating the iterations. From the first column of Figure~\ref{fig:sinkhorn}, we can see that the original cost matrix is chaotic, with many irrelevant regions contributing, which partially demonstrates that simply averaging all these contributions, as done in VSE, will lead to inferior results. The initial transport plan is extremely sparse and meaningless under a small $\lambda$, but after row-wise and column-wise projections, the transport plan becomes diverse and meaningful. As the iterations progress, $\mOmega^{(t)}$ gradually eliminates the disturbance from meaningless regions and concentrates more on relevant regions. Eventually, all other irrelevant regions are ignored, resulting in an optimal plan that aligns the significant fragmented pairs.

\newcommand{\beginsupplementB}{%
    \setcounter{table}{0}
    \renewcommand{\thetable}{B\arabic{table}}%
    \setcounter{figure}{0}
    \renewcommand{\thefigure}{B\arabic{figure}}%
    \setcounter{algorithm}{0}
    \renewcommand{\thealgorithm}{B\arabic{algorithm}}%
    \setcounter{section}{0}
    \renewcommand{\thesection}{B\arabic{section}}%
}
\beginsupplementB
\section{Supplements to Experiments}
\begin{figure*}[htbp]
    \centering
    \captionsetup{aboveskip=0pt, belowskip=0pt}

    {\centering
    \Large\textit{\textbf{Query}: Two men standing on the back of a cart pulled by horses.}\par
    }

    \begin{subfigure}[b]{0.497\linewidth}
        \centering
        \includegraphics[width=\linewidth]{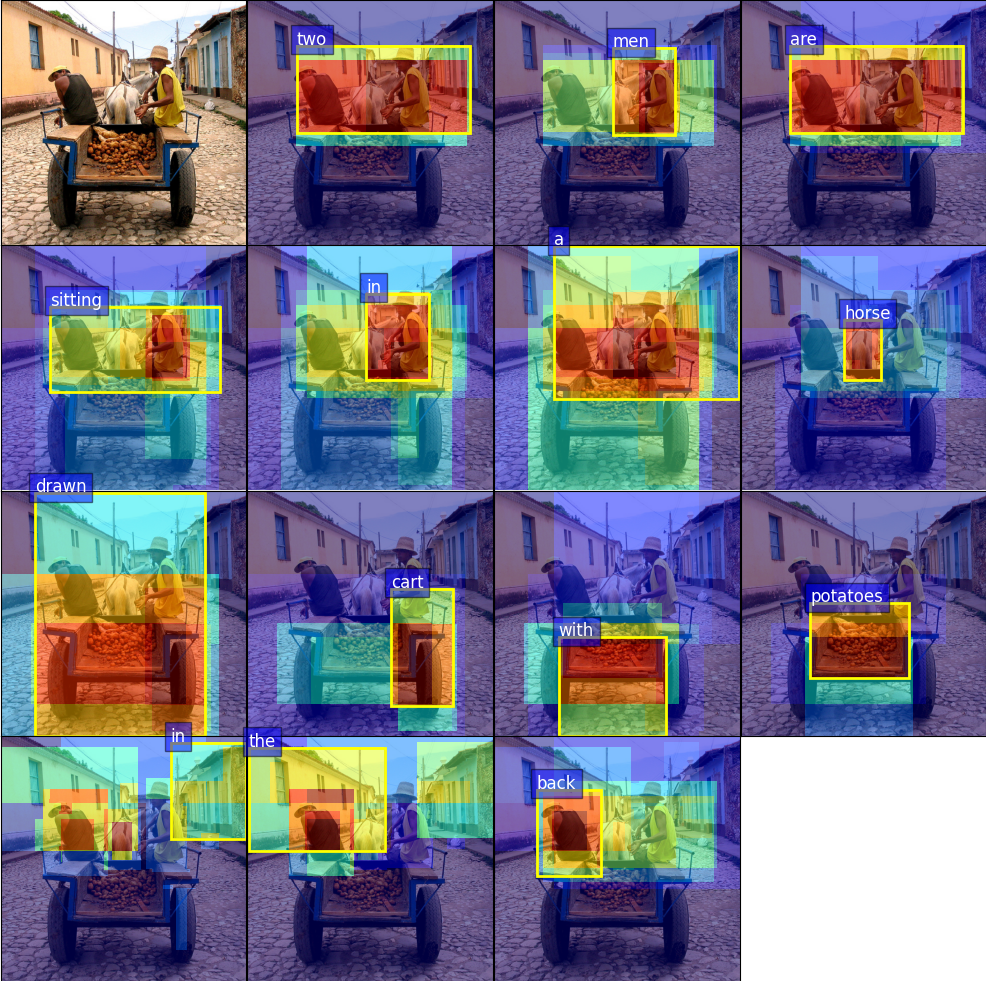}
        \label{fig:men_region}
    \end{subfigure}
    \begin{subfigure}[b]{0.497\linewidth}
        \centering
        \includegraphics[width=\linewidth]{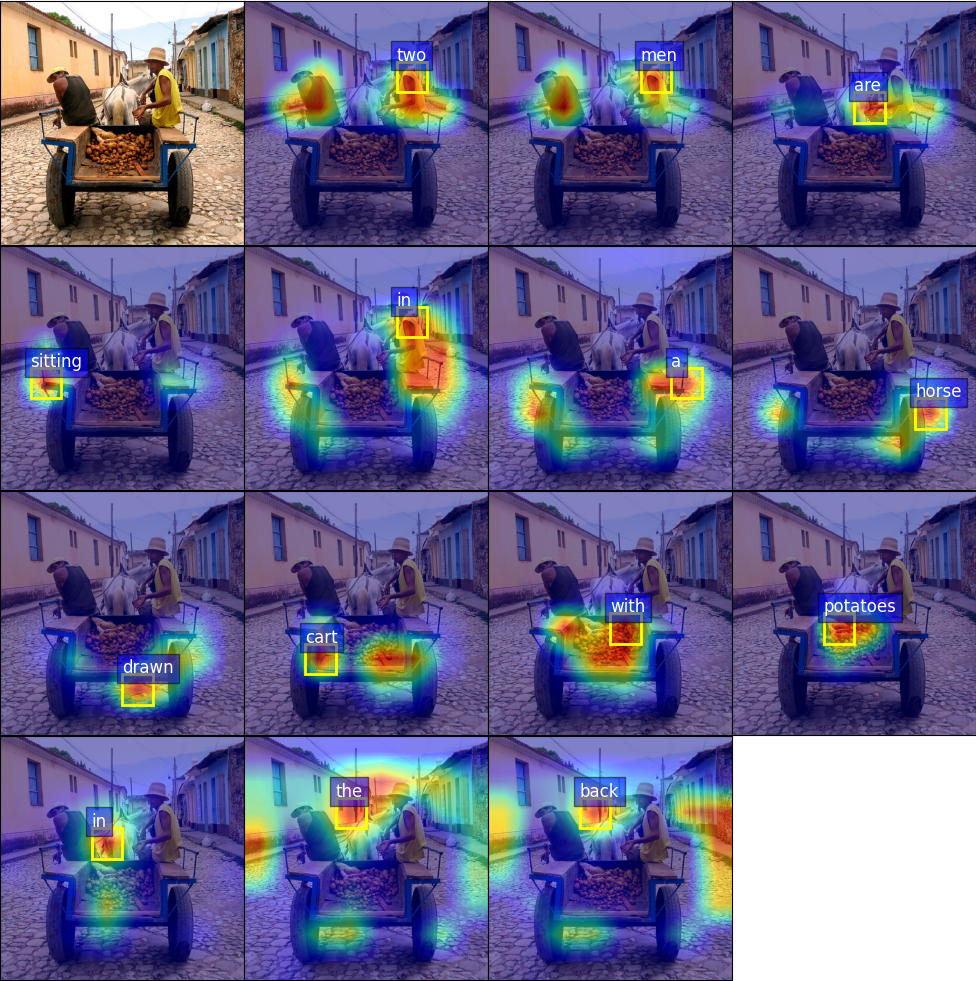}
        \label{fig:men_grid}
    \end{subfigure}

    \caption{
    An illustrative example of image-text matching showcasing the column-wise transport plan generated by our OMIT variants. 
    Left: region-based OMIT; Right: grid-based OMIT. 
    \cmmnt{This example provides clear evidence that OMIT excels in aligning granular semantics. 
    Compared to the region-based variant, the grid-based OMIT encounters challenges in accurately inferring contextual information. 
    Specifically, the model erroneously interprets the word "back" as referring to the background instead of the back of the cart.}
    }
    \label{fig:men_combined}
\end{figure*}

\begin{figure*}[htbp]
    \centering
    \captionsetup{aboveskip=0pt, belowskip=0pt}

    {\centering
    \Large\textit{\textbf{Query}: Two girls with one girl carrying another girl on her back.}\par
    }

    \begin{subfigure}[b]{0.497\linewidth}
        \centering
        \includegraphics[width=\linewidth]{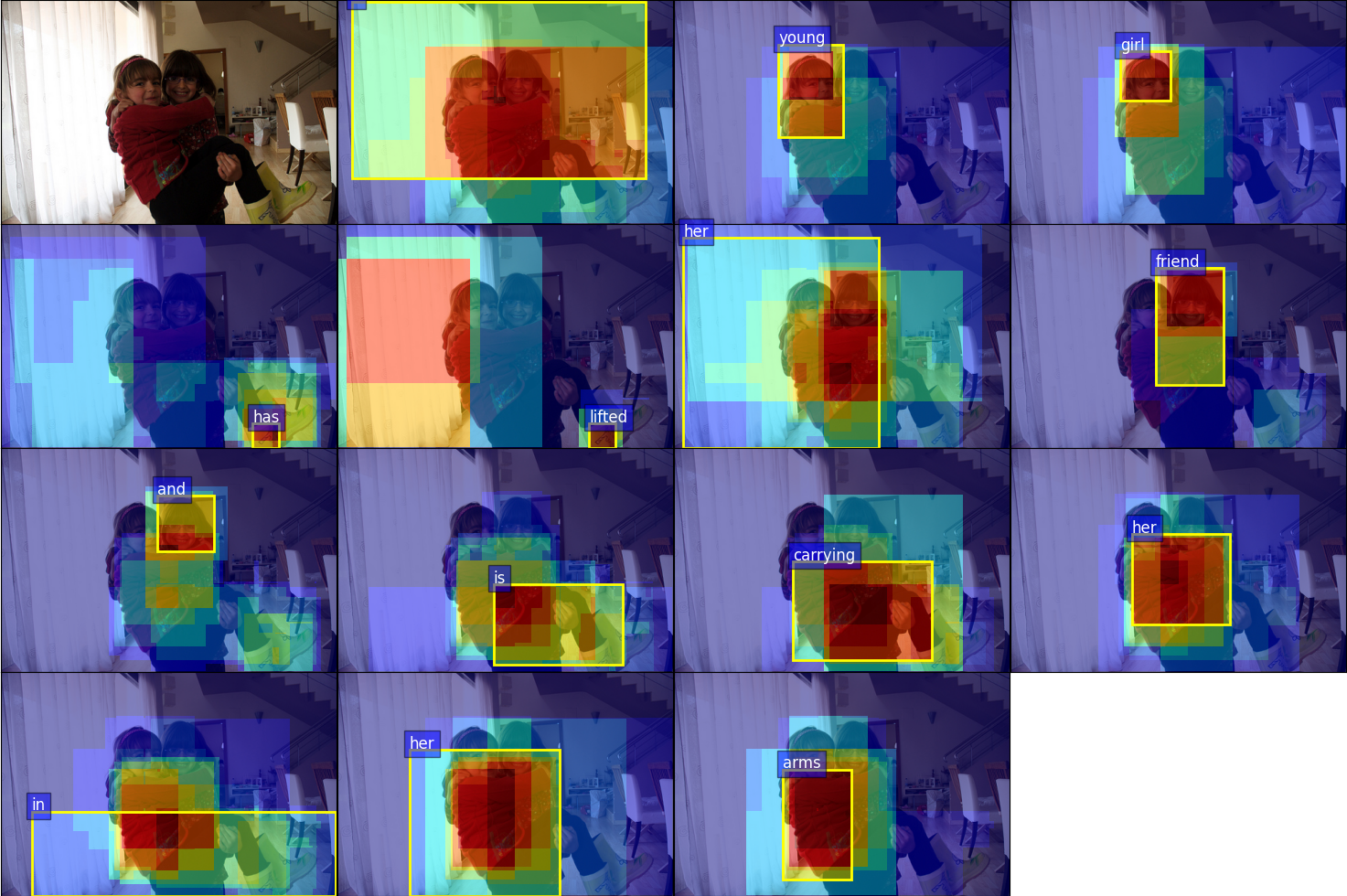}
        \label{fig:carrying_region}
    \end{subfigure}
    \begin{subfigure}[b]{0.497\linewidth}
        \centering
        \includegraphics[width=\linewidth]{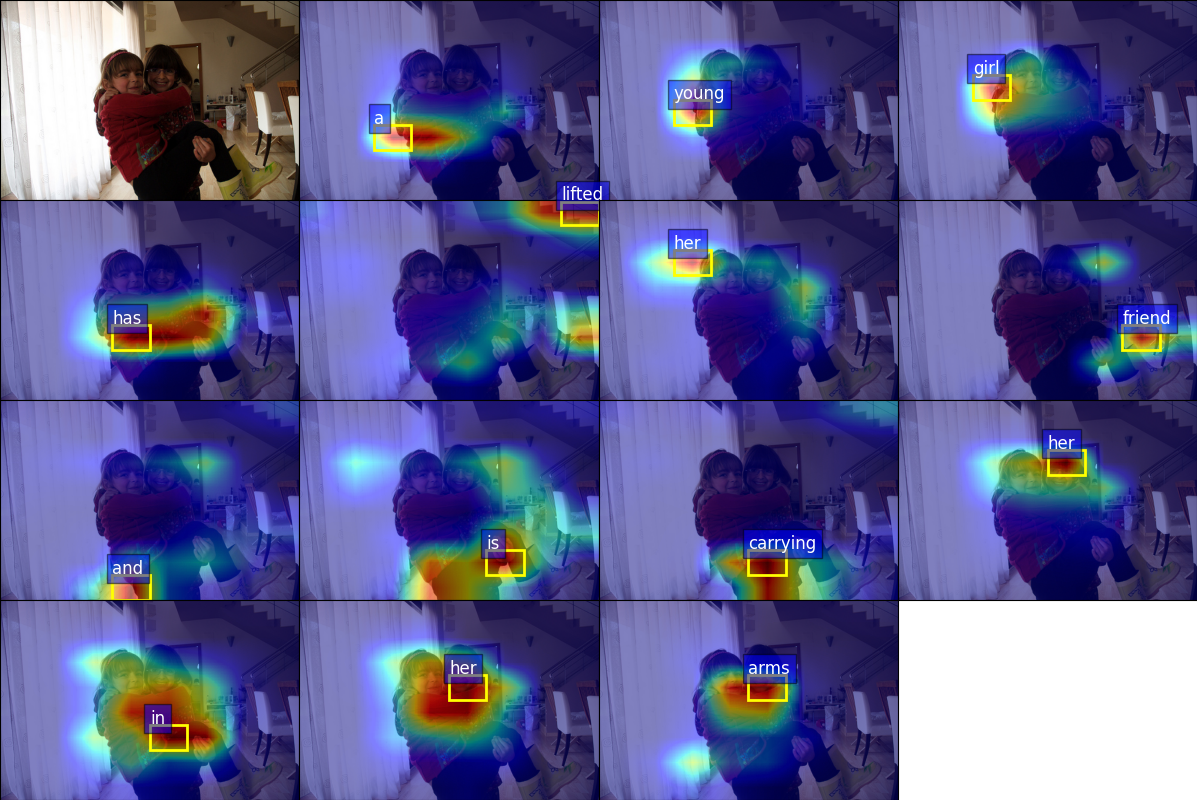}
        \label{fig:carrying_grid}
    \end{subfigure}

    \caption{
    An illustrative example of image-text matching demonstrating the column-wise transport plan generated by our OMIT variants. 
    Left: region-based OMIT; Right: grid-based OMIT. 
    \cmmnt{In this instance, our OMIT encounters difficulties in accurately distinguishing between two girls and mistakenly identifies the girls performing the lifting action as the girl being lifted. 
    The grid-based variant further shows that the model struggles to grasp the deeper meaning and context underlying the image-text relationship.}
    }
    \label{fig:carrying_combined}
\end{figure*}

\subsection{More Ablation Studies}

\mypara{Compared With Different Margin Initialization Strategies.} In light of the observed disparity in marginal conditions between $\mathcal{S}_{CAM}$ and $\mathcal{S}_{OT}$, we delve into the performance analysis of OMIT under various margin initialization strategies. We consider four different types of margins motivated by current literature: "Uni"~\cite{peyre2019computational}, "Intra"~\cite{liu2019focus}, "Inter"~\cite{zou2022tokenflow} and "Norm"~\cite{yokoi2020word}. In our main text, we have already adopted the uniform margins denoted as "Uni". Here, we provide a brief explanation of each type of margin initialization:
\begin{itemize}
    \item "Uni":~Uniform margins used in our main text.
    \item "Intra":~Margins initialized based on the proximity of each fragment embedding to their respective intra-modal global embedding.
    \item "Inter":~Margins initialized based on the proximity of each fragment embedding to their inter-modal global embedding.
    \item "Uni":~Margins initialized based on the relative normalization value of the embeddings.
\end{itemize}
 Mathematically, these margins are defined as follows:
\allowdisplaybreaks[4]
\begin{align}
    \text{\textbf{Uni:}}~ \valpha_i=\frac{1}{K_{m}}&,~\vbeta_j=\frac{1}{L_{n}} \notag \\
    \text{\textbf{Intra:}}~ \valpha_i=\frac{\mathop{exp}(\overline{\vv}^{\top} \vv_i/\tau)}{\sum_{i}^{K_{m}}\mathop{exp}(\overline{\vv}^{\top} \vv_i/\tau)}&,~\vbeta_j=\frac{\mathop{exp}(\overline{\vt}^{\top} \vt_j/\tau)}{\sum_{j}^{L_{n}}\mathop{exp}(\overline{\vt}^{\top} \vt_j/\tau)} \notag  \\
    \text{\textbf{Inter:}}~ \valpha_i=\frac{\mathop{exp}(\overline{\vt}^{\top} \vv_i/\tau)}{\sum_{i}^{K_{m}}\mathop{exp}(\overline{\vt}^{\top} \vv_i/\tau)}&,~\vbeta_j=\frac{\mathop{exp}(\overline{\vv}^{\top} \vt_j/\tau)}{\sum_{j}^{L_{n}}\mathop{exp}(\overline{\vv}^{\top} \vt_j/\tau)} \notag  \\
    \text{\textbf{Norm:}}~ \valpha_i = \frac{ \| \widetilde{\vv_i} \|_2}{\sum_{i}^{K_{m}}{\|\widetilde{\vv_i} \|_2}},~ \vbeta_j &= \frac{\|\widetilde{\vt_j}\|_2}{\sum_{i}^{L_{n}}{\|\widetilde{\vt_j}\|_2 }}
    \centering
    \label{eq:sup_margins}
\end{align}

Here, $\vv_i$ and $\vt_j$ denote one of the fragment embeddings. $\widetilde{\vv_i}$ and $\widetilde{\vt_j}$ are fragment embeddings before length normalization, they satisfies $\vv_i = \frac{\widetilde{\vv_i}}{\|\widetilde{\vv_i}\|}$ and $\vt_j = \frac{\widetilde{\vt_j}}{\|\widetilde{\vt_j}\|}$. $\tau$ is the temperature parameter. Based on empirical observations, we set $\tau=1$ in Equation \ref{eq:sup_margins} since we have found that using a smaller value for $\tau$ leads to degraded performance.

The results in Table~\ref{tab:cmpr_mrg} indicate that the choice of different marginal types has limited impact on the performance of the Cross-modal Mover's Distance (CMD), highlighting its robustness within certain limits. Specifically, among the various marginal types examined, the "Intra" variant consistently performs the worst in both benchmarks. This observation suggests that relying solely on the relative proximity within a modality is an unreliable approach for calculating CMD. Conversely, the uniform margins exhibit competitive performance compared to the "Inter" and "Norm" variants. We attribute this success to the assumption of equal contributions from intra-modal fragments, which serves as a significant prior in CMD. 

\begin{table}[tbp]
    \centering
    \small
    {
        \tabcolsep 4 pt
	    \begin{tabular}{@{\;} l c cccccc@{\;}}
	        \addlinespace
	        \toprule
	        \multicolumn{1}{c}{\multirow{2}{*}{\symtext{TYPE}}} &  \multicolumn{3}{c}{\textsc{img}~$\rightarrow$~\textsc{text}} &      \multicolumn{3}{c}{\textsc{text}~$\rightarrow$~\textsc{img}} & \multicolumn{1}{c}{\multirow{2}{*}{\symtext{rsum}$\uparrow$}} \\ 
	        \cmidrule(lr){2-4} \cmidrule(lr){5-7}
	        {} & \symtext{r}{@1} & \symtext{r}{@5} & \symtext{r}{@10} & \symtext{r}{@1} & \symtext{r}{@5} & \symtext{r}{@10} & \\
                \midrule
                \multicolumn{8}{@{\;}l}{\bf {ResNet-152 \cite{he2016deep} $+$ BiGRU~\cite{chung2014empirical}}} \\[1.5pt]
                \rowcolor{Gray} Uni & 67.0 & 90.3 & 95.6 & 51.8 & 79.4 & 87.1 & \first{471.2} \\
                Intra & 65.4 & 88.6 & 93.9 & 51.5 & 78.6 & 86.2 & 464.3 \\
                Inter & 67.7 & 89.3 & 94.8 & 51.9 & 79.5 & 86.9 & 470.1 \\
                Norm & 68.2	& 88.8 & 94.2 & 52.2 & 79.8	& 87.4 & 470.6 \\
                \midrule
	        \multicolumn{8}{@{\;}l}{\bf {BUTD\cite{anderson2018bottom} $+$ BiGRU~\cite{chung2014empirical}}} \\[3pt]
		      \rowcolor{Gray} Uni & 78.8 & 95.4 & 98.0 & 60.6 & 85.1 & 91.1 & 509.0 \\
                Intra & 78.3 & 95.4 & 97.8 & 59.7 & 85.1 & 90.8 & 507.1 \\
                Inter & 78.9 & 95.4 & 98.3 & 59.9 & 84.8 & 91.0 & 508.3 \\
                Norm & 80.0	& 95.9 & 97.4 & 59.6 & 85.2 & 91.0 & \first{509.1} \\
	       \bottomrule
	    \end{tabular}
	}
    \caption{Comparisons of OMIT variants with different marginal conditions on the Flickr30K test set.}
    \label{tab:cmpr_mrg}
\end{table}


\subsection{More Visualizations}

Figures~\ref{fig:men_combined}-\ref{fig:carrying_combined} illustrate the comprehensive alignments achieved by OMIT when utilizing two sets of backbones: BUTD+BiGRU and ResNet-152+BiGRU. In-depth analysis of these figures reveals that ostensive semantics, such as "vehicle," "horse," and "girl," can be effectively represented using only a few sub-regions or pixels. However, the visualization of abstract concepts, function words, and relational words like "back," "of," and "lifted" presents challenges in visual representation, leading to figures that display meaningless chunks or even provide incorrect indications. Furthermore, in Figure~\ref{fig:men_combined} and Figure~\ref{fig:carrying_combined}, OMIT encounters difficulties in distinguishing between the two girls, incorrectly identifying the girls lifting as the girl being lifted. These observations highlight the inherent limitation of OMIT, which primarily models co-occurrences but lacks the cognitive and reasoning capabilities necessary for accurate understanding. We speculate that incorporating the high-order relations may help overcome these limitations and eagerly anticipate further exploration in this direction.

\end{appendix}

\clearpage
{\small
\bibliographystyle{ieeenat_fullname}
\bibliography{bibs/aaai2025}
}

\end{document}